\documentclass{iopart}
\usepackage[colorlinks=true]{hyperref}
\usepackage{graphicx}
\usepackage{iopams}
\newcommand{\matt}{\mathrm{matt}} 
\newcommand{\bg}{\mathrm{bg}} 
\newcommand{\dt}{\Delta t}       
\newcommand{\dr}{\Delta r}       

\begin{document}
\title{Simulating nonlinear neutrino flavor evolution}

\author{H Duan$^1$, G M Fuller$^2$ and J Carlson$^3$}

\address{$^1$ Institute for Nuclear Theory, %
University of Washington, %
Seattle, WA 98195, USA}
\address{$^2$ Department of Physics, %
University of California, San Diego, %
La Jolla, CA 92093, USA}
\address{$^3$ Theoretical Division, Los Alamos National Laboratory, %
Los Alamos, NM 87545, USA}

\ead{\mailto{hduan@phys.washington.edu}, 
\mailto{gfuller@ucsd.edu}, \mailto{carlson@lanl.gov}}

\begin{abstract}
We discuss a new kind of astrophysical transport problem: the coherent
evolution of neutrino flavor
in core collapse supernovae. Solution of this problem requires
a numerical approach which can simulate accurately the quantum
mechanical coupling of intersecting neutrino trajectories and the
associated nonlinearity which characterizes neutrino flavor conversion.
We describe here the two codes developed to attack this problem. We also
describe the surprising phenomena revealed by these numerical
calculations. Chief among these is that the nonlinearities in the
problem can engineer neutrino flavor transformation which is
dramatically different than in standard Mikheyev-Smirnov-Wolfenstein
treatments. This happens even though the neutrino mass-squared
differences are measured to be small, and even when neutrino
self-coupling is sub-dominant. Our numerical work has revealed potential
signatures which, if detected in the neutrino burst from a Galactic core
collapse event, could reveal heretofore unmeasurable properties of the
neutrinos, such as the mass hierarchy and vacuum mixing angle 
$\theta_{13}$.   
\end{abstract}

\maketitle

\section{Introduction%
\label{sec:introduction}}

The ghost-like neutrinos ($\nu_e$, $\bar\nu_e$, $\nu_\mu$,
$\bar\nu_\mu$, $\nu_\tau$ and $\bar\nu_\tau$)
are chargeless, spin--$\case{1}{2}$ particles notorious 
for their weak interactions. 
The revelation by recent observations and experiments of non-zero
neutrino rest masses and vacuum neutrino flavor mixing  has forced the
astrophysics community to confront a vexing problem: the nonlinear
evolution of neutrino flavor in astrophysical environments where there
are appreciable neutrino fluxes.   Solution of this problem could be
important because many environments associated with compact objects and
the very early universe are energetically and in other ways dominated by
neutrinos and their interactions. Perhaps the most impressive example of
this is in core collapse supernovae, where the collapse of a
highly-evolved core supported by relativistically-degenerate electrons
to a cold neutron star releases $\sim 10\%$ of the core's {\it rest
mass} in neutrinos of all kinds (see, e.g., \cite{Woosley:2006ie}
for an introductory review and \cite{Cardall:2007dy} for a discussion
of general supernova neutrino physics issue).

Neutrinos in this environment are thought to play a pivotal role in
nearly every aspect of supernova physics, from the explosion mechanism
itself, at fairly early times after the collapse of the core, to setting
the conditions for the synthesis of heavy elements at later times. By
virtue of their weak interactions and tenuous coupling to ordinary
matter, neutrinos can transport energy, entropy, and lepton number
through very dense matter that other particles might not be able to
penetrate. And neutrinos can more than make up for their feeble
individual interactions with huge numbers.

The astrophysics community has long recognized the important role that
neutrinos play in core collapse supernova explosions. From the earliest
work of Colgate, Wilson, and others 
\cite{Colgate:1966aa,Bowers:1982aa,Bruenn:1985aa,Myra:1987aa} 
to the most recent and
most sophisticated numerical simulations (e.g.,
\cite{Mezzacappa:2000jb,Fryer:2003jj,Walder:2004ym,Kifonidis:2005yj,Scheck:2006rw,Blondin:2006yw}),
the key problems have been to characterize the
transport of neutrinos in and above the neutron star core and to
calculate the energy spectra and fluxes of each neutrino species.
Solving these problems has required not only front-line coupled
multi-dimensional hydrodynamics and Boltzmann neutrino transport
computations but also understanding physics which is not readily
accessible in the laboratory, such as the equation of state of hot,
neutron-rich nuclear matter. 

However, this game changes somewhat in light of the {\it experimental
fact} of neutrino flavor transformation. It has been shown that
$\nu_e$ and $\bar\nu_e$, neutrinos and antineutrinos with
electron flavor (i.e., neutrinos associated with electrons and positrons
in weak interactions), can be transformed into neutrinos and antineutrinos
with other flavors, $\nu_\mu$, $\nu_\tau$, $\bar\nu_\mu$ and $\bar\nu_\tau$
(see, e.g., \cite{PDBook} for a review).
 Now both goals of calculating
neutrino transport and finding the emergent neutrino energy spectra and
fluxes may not be attainable without an adequate treatment of
medium-affected neutrino flavor transformation. As we shall see below,
in many circumstances adding on a flavor evolution solver would greatly
complicate neutrino transport calculations. We can identify two broad
regimes in this problem: 
(1) the high density partially coherent or  non-coherent regime, where
the neutrino transport mean free paths are comparable to or smaller than
neutrino flavor oscillation lengths or Mikkheyev-Smirnov-Wolfenstein (MSW) 
\cite{Wolfenstein:1977ue,Wolfenstein:1979ni,Mikheyev:1985aa} resonance widths;
(2) the coherent regime, where neutrinos are essentially
free streaming on scales relevant for neutrino flavor 
transformation. 
The first regime can be very difficult to treat and 
requires a quantum kinetic approach
(e.g., \cite{Sigl:1992fn,Abazajian:2001nj,Strack:2005ux,Hidaka:2006sg,Hidaka:2007se,Cardall:2007zw}). 
However, it is generally 
thought that active-active neutrino flavor transformation is negligible 
in this regime. This prejudice is motivated by the fact that the
relevant active-active channel neutrino mass-squared differences are small,
the matter density is high in this regime \cite{Fuller:1987aa,Sigl:1992fn},
and collective neutrino flavor transformation is suppressed in
very dense neutrino gases \cite{Duan:2005cp}.
Our numerical approach concentrates on the second regime.

The coherent neutrino flavor evolution problem differs in two essential
aspects from conventional transport problems. First, of course, is
quantum mechanical coherence itself. Second is neutrino self-coupling. 

Coherence dictates that we follow the development in time and/or space
of the complex amplitudes which describe the flavor states of neutrinos.
This must be done with a Schr\"odinger-like equation. On a given
neutrino trajectory, or world line, with Affine parameter $t$, this is
\begin{equation}
\rmi\frac{\rmd}{\rmd t}|\psi(t)\rangle=
\hat{H}|\psi(t)\rangle.
\label{eq:schroedinger}
\end{equation}
Here the operator $\hat{H}$ is a Hermitian Hamiltonian operator that
contains the vacuum and in-medium generators of neutrino flavor
evolution, while $|\psi\left( t\right)\rangle$ is the ket that
describes the flavor state of a neutrino at point $t$ on this given
trajectory. Neutrinos in supernovae for the most part will be
ultra-relativistic, so it may seem odd that a ``non-relativistic
Schr\"odinger equation'' can give an adequate account of flavor
evolution. It works for two reasons \cite{Halprin:1986pn}: (1) we consider
only neutrino forward scattering interactions and neglect inelastic
scattering; and (2) in the ultra-relativistic limit the information in
the four-component Dirac spinor describing the neutrino's flavor state
is mostly contained in two components, the two ``large''
components.

\begin{figure}[t]
\begin{indented}
\item[]%
\includegraphics*[angle=-90,width=0.5\textwidth,keepaspectratio]%
{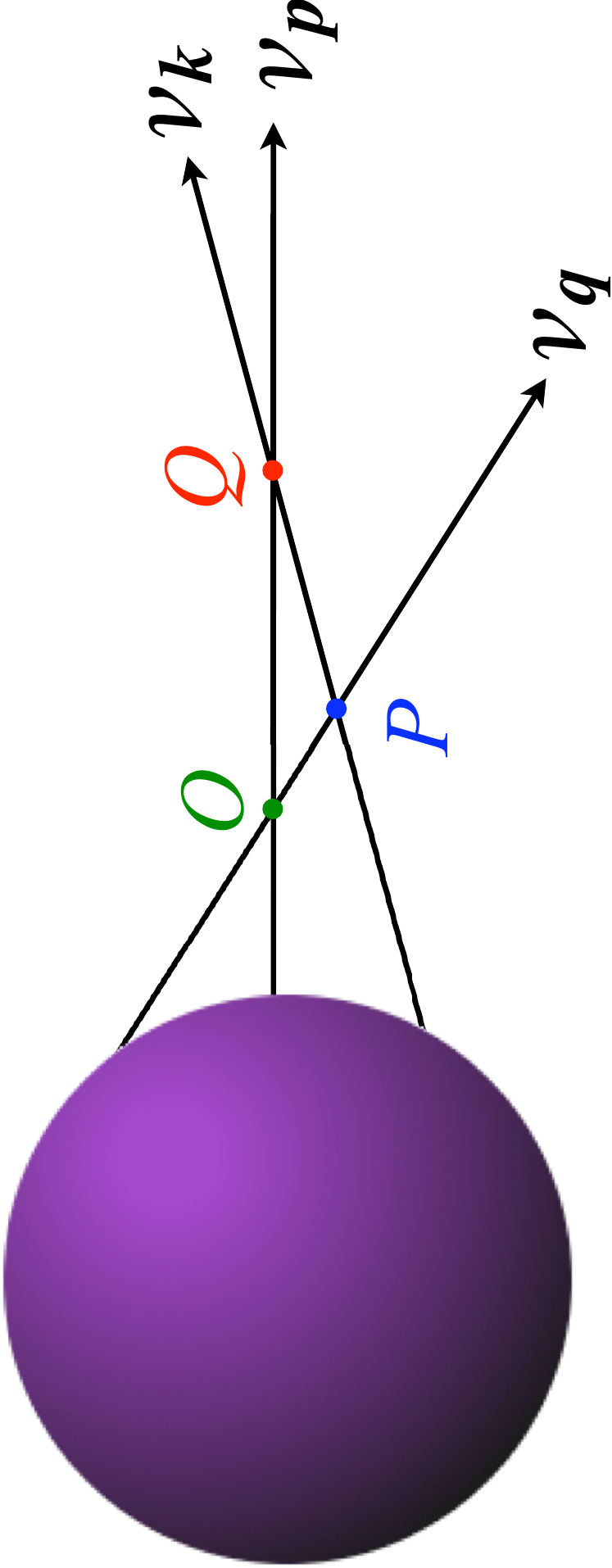}
\end{indented}
\caption{\label{fig:entanglement}
Illustration of the quantum coupling problem for
neutrinos on different, but intersecting trajectories. The
world lines for three
neutrino beams, $\nu_\bi{k}$, $\nu_\bi{p}$ and $\nu_\bi{q}$,
intersect as shown at points $O$, $P$ and $Q$.
Neutrino-neutrino forward scattering at these intersection points
will quantum mechanically couple 
the flavor evolution histories of these neutrinos.}
\end{figure}

A serious complication in following neutrino flavor transformation in
supernovae stems from neutrino-neutrino forward scattering contributions
to $\hat{H}$. These render the problem nonlinear, since the interactions
which dictate flavor transformation amplitudes are themselves dependent
on the neutrino flavor states. This seemingly innocuous statement masks
a kind of geometric coupling which, to our knowledge, is unique in
astrophysical transport problems. Neutrino-neutrino scattering could
quantum mechanically couple the flavor evolution on all intersecting
trajectories. This is illustrated in Figure \ref{fig:entanglement}. 
Following the flavor
evolution of neutrino $\nu_\bi{p}$ would entail knowing the flavor states of
all neutrinos on which it forward-scattered. That might include, for
example, neutrinos $\nu_\bi{k}$ and $\nu_\bi{q}$ 
propagating on world lines which
intersect $\nu_\bi{p}$'s trajectory at points {\it Q} and {\it O},
respectively. Note, however, that $\nu_\bi{k}$'s and $\nu_\bi{q}$'s flavor
histories are not independent, since they may have undergone a forward
scattering event at point {\it P}. Obviously, with a large number of
neutrinos this quickly can become a daunting problem!

To simplify the problem we adopt the spherically
symmetric ``neutrino bulb'' model (hereafter the ``bulb model'')
\cite{Duan:2006an}.
In this model neutrinos are emitted half-isotropically
from a spherical ``neutrino sphere'' (essentially the hot
neutron star surface) with energy spectra that are initially of black
body form. We adopt a smooth matter density profile $\rho$ which
depends only on radius $r$, 
the distance from the center of the neutron star.
At late times, close to the neutron star 
where matter tends to be isothermal,
this matter profile falls exponentially.
Further out where the envelope may be
closer to constant entropy in character,
the density profile is  $\rho\sim r^{-3}$. 
Because of the assumed spherical symmetry,
neutrino trajectories with the same emission angle
$\vartheta_0$ become equivalent,
and neutrino modes emitted in the same flavor and energy state
and with the same emission angle have identical flavor evolution
history \cite{Duan:2006an}.
Here the emission angle $\vartheta_0$ is the angle
between the neutrino propagation direction and the vector normal
to the neutrino sphere at the emission point.

In the bulb model, all neutrino evolution above the neutrino sphere is assumed
to be coherent and essentially free-streaming in character. This will be
a good approximation for the conditions expected to obtain a few seconds
after core bounce, and may be useful for earlier times in some
cases. We bin neutrino modes emitted in each flavor by both the emission
angle and the energy of the neutrino mode. Because the flavor histories
of all neutrino bins are coupled, we solve self-consistently
for the flavor evolution
of all neutrino modes simultaneously. The flavor evolution of
each neutrino mode is described by an
equation with the form of Equation (\ref{eq:schroedinger}).
Each neutrino trajectory has an individual Affine parameter
which can be expressed in terms of radius $r$.

There are a variety of physical processes which can affect neutrino flavor 
evolution, either by complicating the coherent calculation 
procedure outlined above, or by introducing decoherent effects. As an example 
of the former, the density profile in the core collapse supernova environment 
can be quite dependent on time, with shocks and multi-dimesional effects of 
paramount importance in some epochs (e.g., 
\cite{Schirato:2002tg,Tomas:2004gr}). 
In the latter case, matter density fluctuations associated
with shocks, sound waves, and turbulence can lead to neutrino flavor 
depolarization and decoherence
\cite{Sawyer:1990tw,Loreti:1995ae,Friedland:2006ta,Kneller:2007kg}.

In our numerical work we have chosen to concentrate on the simplest
environments in order to gain insight into the effects of coupled
quantum mechanical flavor evolution. Our rationale is that this physics
will be present always in the environments we treat. The nonlinear
collective phenomena revealed by our calculations can be modified
in real supernovae by the kinds of effects discussed above. However, our
calculations lead us to conclude that most of the {\it qualitative} and
even some of the quantitative results for our simple models are likely
to survive. Even our simplified supernova models present novel numerical
challenges which will have to be faced in any 
comprehensive treatment of the subject.

\section{Solution method%
\label{sec:method}}

Even with the simplified model discussed in Section \ref{sec:introduction},
obtaining a convincing solution to the full, coupled-trajectory
supernova neutrino flavor evolution problem, which we term the 
``multi-angle'' problem, is still a challenging task. So difficult, in
fact, that there had been no attempt to solve it prior to our effort.
This meant that previous work provided no guidance on numerical strategies 
for solving multi-angle neutrino flavor evolution. The only clues to general 
coherent neutrino flavor phenomena in supernovae came from so-called 
``single-angle'' calculations, where trajectories were coupled, but 
the flavor evolution on every trajectory was taken to be the same as that on 
a radially-directed trajectory.
In Section \ref{sec:plan} we outline the plan we used to attack this problem.
Partly because of the lack of previous work, we
have developed two numerical codes more or less 
independently at UCSD and LANL. 
As an aid to debugging and as a device for teasing
apart physics issues from numerical issues, the results from the two
codes were compared for key test problems. This procedure proved
efficacious, in part because sometimes numerical issues and problems
tended to be different in each code. In Sections
\ref{sec:flat-structure}--\ref{sec:nbgroupmm}
we highlight the key implementations of \textit{FLAT},
the computer code developed at UCSD. In Section \ref{sec:bulb} we
discuss some features of \textit{BULB}, the code
developed at LANL.

\subsection{Project plan%
\label{sec:plan}}

Our ultimate goal is to solve for the flavor histories
of neutrinos and antineutrinos
(i.e.\ $\nu_e$, $\nu_\mu$, $\nu_\tau$, $\bar\nu_e$, $\bar\nu_\mu$ 
and $\bar\nu_\tau$) emitted in the bulb model with
realistic energy spectra and traveling on all
physically realizable trajectories. This is what we term the 
``three-flavor multi-angle'' problem. 

In the literature, the neutrino oscillation problem with three flavors
is usually approximated as two two-flavor oscillation problems
that occur at different time/distance scales. The rationale for this is that
the neutrino mass-squared differences associated with these two-flavor
oscillation problems are different by more than an order of magnitude.
Previous work on medium-affected neutrino oscillations exclusively adopted
this two-flavor approximation
\cite{Sigl:1992fn,Samuel:1993uw,Kostelecky:1994dt,Qian:1994wh,Pastor:2001iu,Pastor:2002we,Balantekin:2004ug,Fuller:2005ae}.

These studies also adopted  what we termed above 
the ``single-angle'' approximation.
In this approximation flavor evolution of neutrinos with energy $E$
propagating along different neutrino trajectories is
assumed to be the same as that of neutrinos with the same energy $E$
propagating along a representative trajectory,
usually taken to be the radial direction.
The validity of both the two-flavor and single-angle
approximations for supernova neutrino flavor evolution, 
however, remains to be shown. 

Because of the big gap between the ultimate goal we would like to achieve
and the state of art of neutrino oscillation studies, it is clearly 
very risky
to attempt a solution to the full problem in one single step. 
Not only will there
be little theoretical guidance in such a monolithic approach,
but also finding physical interpretations of any novel numerical results
could prove to be problematic. Instead, we have sought
to achieve our goal in four stages.

In the first stage we solve for neutrino flavor transformation
using the two-flavor, single-angle approximation.
This stage overlaps with studies which already exist in the literature.
This procedure should offer a good check on our numerical codes 
as well as on existing theories.

In the second stage we will still assume two neutrino flavors,
but will solve for neutrino and antineutrino flavor evolution 
using a full multi-angle implementation of the bulb model.
The difference between this stage and the first stage should
give us insights on how multi-angle effects may enter into
neutrino flavor transformation and how good the single-angle
approximation might be.

In the third stage we will implement full three-neutrino flavor evolution
but with the single-angle approximation. Comparison of the
results of this stage and those of the first stage should give
us important clues on whether the two-flavor approximation is valid
in collective neutrino oscillations.

In the last stage we will solve the full problem, i.e.,
neutrino oscillations with three neutrino flavors and 
a full multi-angle implementation.
Presumably, the results of the computation in this stage
could be estimated at least crudely from the theories which
employ the two-flavor, single-angle model and
the results of the second and third stages calculations.
 Comparison between the
numerical results in the full three-neutrino calculations
and the theoretical estimates again can be
used to check both the numerical codes and the theories.

\subsection{\textit{FLAT}'s highly modular design%
\label{sec:flat-structure}}

\begin{figure}[t]
\begin{indented}
\item[]%
\includegraphics*[angle=-90,origin=bl,width=0.8
\textwidth,keepaspectratio]%
{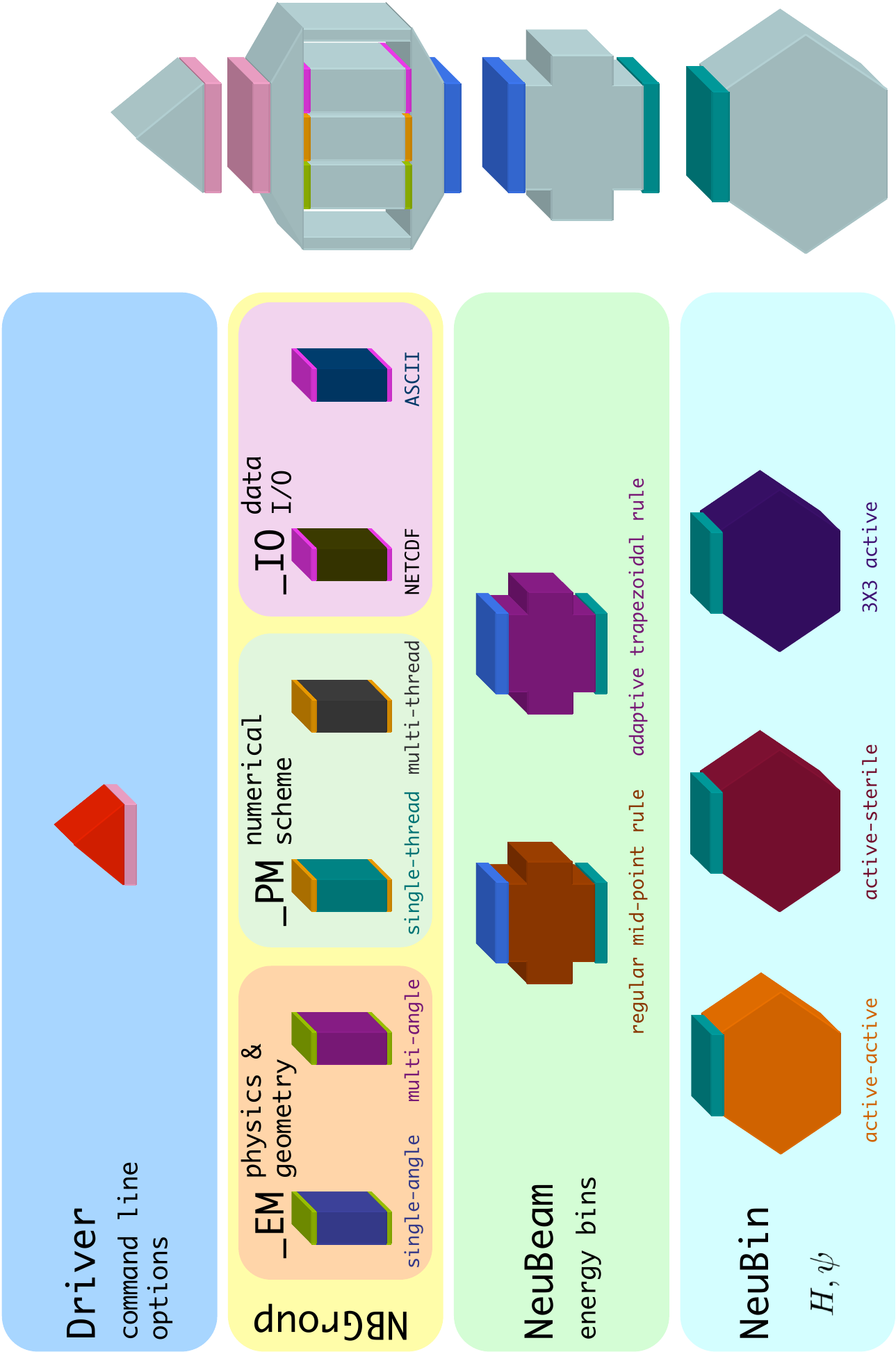}
\end{indented}
\caption{\label{fig:flat}%
Illustration showing the structure of \textit{FLAT}. This code is a
collection of modules. These modules are represented here 
as various building blocks in the left part of
the figure. An executable binary can be compiled by choosing
appropriate modules for a particular problem. This is illustrated in
the right part of the figure. Building blocks with the same geometric
shape and the same top/bottom layer(s), but with different colors,
represent modules
that have the same interface functions but perhaps employ
different physical approximations or numerical algorithms.
Changing a particular physical setting or a numerical
algorithm can be achieved by swapping out the appropriate
module(s) in the code with other module(s) with the same interface functions.}
\end{figure}

As explained above, we plan to
achieve our ultimate goal progressively, essentially by solving
three intermediate problems before attacking the full problem.
These problems differ only in the physical approximations that they employ.
 Written in C++, \textit{FLAT} takes advantage of the
Object-Oriented Programming paradigm and
provides a unique and uniform solution to all these problems
by offering a hierarchical set of problem-specific modules. 
Changing a particular physical setting or a numerical
algorithm can be done by swapping out appropriate module(s).
The structure of \textit{FLAT} is
illustrated in Figure \ref{fig:flat}.

At the lowest level of the hierarchy are the \texttt{NeuBin} modules.
Each of the  \texttt{NeuBin} modules implements two C++ classes,
\texttt{NeuPot} and \texttt{NeuBin}, which correspond
to the Hamiltonian $\hat{H}$ and the flavor state $|\psi\rangle$
in Equation (\ref{eq:schroedinger}), respectively. We note that the
Schr\"{o}dinger Equation (\ref{eq:schroedinger}), as well as the
numerical algorithm that \textit{FLAT} uses to solve this equation, do
not depend
on the specifics of the representations of $\hat{H}$ or $|\psi\rangle$.
Therefore,
the higher-level modules in \textit{FLAT} are independent of the
internal implementation in the \texttt{NeuBin} modules 
and can operate through 
a uniform interface on any \texttt{NeuBin} module.
The \texttt{NeuBin} modules are represented as
pentagonal blocks in Figure \ref{fig:flat}. The 
\texttt{NeuBin} module interfaces are represented by the 
common-color layers on top of the pentagonal blocks.

Two examples of the interface functions of a  \texttt{NeuBin}
class are \texttt{NeuBin::evolve()} and \texttt{NeuBin::diff()}.
Function \texttt{NeuBin::evolve()} evolves the corresponding
flavor state $|\psi\rangle$ residing at Affine parameter value $t$ on a
given trajectory one step further
for given  Hamiltonian $\hat{H}$ and step size $\dt$
 (see Section \ref{sec:neubin}). 
Function \texttt{NeuBin::diff()}
computes the ``difference'' between two flavor states. This
difference is used to estimate numerical errors.

There are two \texttt{NeuBin} modules that have been used
in production runs: \texttt{NeuBin\_F2C} and \texttt{NeuBin\_F3C}.
These modules define the flavor state and Hamiltonian
for two-flavor and three-flavor problems, respectively.
Once the program has been proven to work well using
the two-flavor approximation, we can use the same program
to solve three-flavor problems by swapping out the
 \texttt{NeuBin\_F2C} module with \texttt{NeuBin\_F3C}.

Above the \texttt{NeuBin} modules
are the \texttt{NeuBeam} modules, each of which defines
 a \texttt{NeuBeam} class. A  \texttt{NeuBeam} object contains
an array of \texttt{NeuBin} objects. Each element of this array
corresponds to a flavor
state $|\psi(E)\rangle$ for a particular energy bin 
$[E-\case{1}{2}\Delta E,E+\case{1}{2}\Delta E]$. 
All \texttt{NeuBeam} modules share
the same interface through which higher-level modules can operate
\texttt{NeuBeam} objects. 

\texttt{NeuBeam::sum()} is one of the important interface functions 
for the \texttt{NeuBeam} modules. This function does a numerical integration
over all energy bins. This summation
represents an important step in producing the neutrino self-coupling
Hamiltonian $\hat{H}_{\nu\nu}$ (see Section \ref{sec:bulb}).
The algorithm used in this numerical
integration is specific to each module and, again, is hidden from
higher-level modules. 

The \texttt{NeuBeam} module used in production runs is 
called \texttt{NeuBeam\_SS}. In this module 
\texttt{NeuBeam::sum()} employs a simple midpoint rule
for numerical integration.
If a more sophisticated algorithm is required, a new \texttt{NeuBeam} module
can be created. The
\texttt{NeuBeam\_SS} module can be swapped out with
this new module without
modification to the remaining components of \textit{FLAT}.
The corresponding upgrade of the code is done automatically. 

On top of the \texttt{NeuBeam} modules are the \texttt{NBGroup} modules.
There are three kinds of \texttt{NBGroup} modules:
\texttt{NBGroup\_EM}, \texttt{NBGroup\_IO}, and \texttt{NBGroup\_PM}.
These modules specify the physical environment, 
the input/output (I/O) methods, and the numerical
algorithm to solve differential equations, respectively.
These three types of modules share the same core data,
i.e., the current flavor states of all neutrinos, and,
at the same time, perform nearly independent tasks.
This special relation among \texttt{NBGroup} modules is
realized through the inheritance of classes and virtual
functions. An \texttt{NBGroup} class is defined first. This class
contains all the data that are used by at least two kinds of 
\texttt{NBGroup} modules.
This class also declares a variety of virtual functions such
as \texttt{NBGroup::set\_bg()}, \texttt{NBGroup::check()}
 and  \texttt{NBGroup::evolve()}.
Each of these virtual functions is defined in 
a child class of \texttt{NBGroup}, but may be called by
other child classes.

Each of the \texttt{NBGroup\_EM} modules defines an
\texttt{NBGroup\_EM} class as a child class of \texttt{NBGroup}. 
A key interface function of
these modules is \texttt{NBGroup\_EM::set\_bg()} which,
by definition of a virtual function, is the ``real function'' 
to be executed when
virtual function \texttt{NBGroup::set\_bg()} is called.
\texttt{NBGroup\_EM::set\_bg()} calculates $\hat{H}_\mathrm{matt}$,
the ordinary matter background contribution to the
total Hamiltonian $\hat{H}$.
\texttt{NBGroup\_EM::set\_bg()} also calculates $\hat{H}_{\nu\nu}$,
the background neutrino medium contribution to $\hat{H}$.

There are two \texttt{NBGroup\_EM} modules used in production runs:
\texttt{NBGroup\_PNS1D} and \texttt{NBGroup\_PNS2D}.  
In \texttt{NBGroup\_PNS1D} the single-angle approximation
is implemented, while in \texttt{NBGroup\_PNS2D} 
the multi-angle implementation is employed.

Each of the \texttt{NBGroup\_IO} modules defines an
\texttt{NBGroup\_IO} class as a child class of \texttt{NBGroup}. 
An \texttt{NBGroup\_IO} class defines \texttt{NBGroup\_IO::check()},
which is the real function corresponding to \texttt{NBGroup::check()}.
\texttt{NBGroup\_IO::check()} performs a variety of critical checks.
For example,  if necessary, 
it can save a snapshot of the current flavor states
of all the neutrino modes, e.g., at specified 
time intervals. 
Also, \texttt{NBGroup\_IO::check()} can set a stop flag
if the specified time allocation is used up.

The \texttt{NBGroup\_IO} modules used
in production runs are \texttt{NBGroup\_NC} and \texttt{NBGroup\_NC\_SMP}.
Module \texttt{NBGroup\_NC}  reads and writes data in the NetCDF format
\cite{Rew:1990aa}, a cross-platform binary
data format. Module \texttt{NBGroup\_NC\_SMP} is similar
to \texttt{NBGroup\_NC}, except that it handles I/O in a
different POSIX thread (Pthread) \cite{Pthread:2004}.
As a result, \texttt{NBGroup\_NC\_SMP} is able to complete
\texttt{NBGroup\_IO::check()}
function calls promptly. This helps to augment the overall code
performance, 
because the CPU can continue to do computation 
during the time in which it would otherwise be idle
waiting for slow I/O data transfers to complete.

Each of the \texttt{NBGroup\_PM} modules
defines an \texttt{NBGroup\_PM} class
as a child class of \texttt{NBGroup}. 
An \texttt{NBGroup\_PM} class 
 defines a member function \texttt{NBGroup\_PM::evolve()},
the real function corresponding to \texttt{NBGroup::evolve()}.
Once called, \texttt{NBGroup\_PM::evolve()} can continuously
evolve neutrino flavor states over radial distance
until stop flags are encountered.

The \texttt{NBGroup\_PM} modules used in production runs are
\texttt{NBGroup\_MM} and \texttt{NBGroup\_MM\_SMP}. 
In the \texttt{NBGroup\_MM} module, \texttt{NBGroup\_PM::evolve()}
employs an adaptive step-size algorithm to solve 
Equation (\ref{eq:schroedinger}) for each neutrino trajectory
(see Section \ref{sec:nbgroupmm}). 
During each sub-step of the algorithm's execution, 
\texttt{NBGroup\_PM::evolve()}
calls \texttt{NBGroup::set\_bg()} to set 
$\hat{H}_\mathrm{matt}+\hat{H}_{\nu\nu}$.
At the end of each whole step in the algorithm's execution,
\texttt{NBGroup\_PM::evolve()} calls
\texttt{NBGroup::check()} 
to check and see if any I/O needs to be done and if the
program needs to be terminated.
\texttt{NBGroup\_MM\_SMP} implements a multi-threaded version
of the algorithm used in \texttt{NBGroup\_MM}. \textit{FLAT}
automatically uses the multi-CPU/core feature
of a modern workstation when the \texttt{NBGroup\_MM} module
is replaced by \texttt{NBGroup\_MM\_SMP}.

At the highest level, a \texttt{Driver} module defines a 
class designated as \texttt{Neutrinos}. This class inherits the
\texttt{NBGroup\_EM},
\texttt{NBGroup\_PM} and \texttt{NBGroup\_IO} classes.
In the \texttt{Driver} module, \texttt{Neutrinos::evolve()} 
(which is actually \texttt{NBGroup\_PM::evolve()}) is 
called to solve the neutrino flavor transformation problem.
\texttt{Driver} also handles command line options. These might include, for
example, setting the time allocation and/or the final radius.

\textit{FLAT}'s highly modular framework endows it with
great extensibility. This  aids program development
as well as helps with the exploration of neutrino flavor
conversion physics. 
With this modular architecture it is straightforward to pick
the simplest numerical algorithm to attack a given problem.
Additional or different modules can be added easily if more 
sophisticated algorithms
are necessary, or if the intention is to treat different problems.
%
%
When following the project plan outlined in Section \ref{sec:plan},
the modular framework of \textit{FLAT}  enables us to minimize  redundancy
in code writing and maximize usage of existing code.
It also maintains consistency when we progress from one
stage to the next and, therefore, isolates the differences
in various physics approximations when we compare
the results obtained in different stages.

\subsection{\texttt{NeuBin} modules in \textit{FLAT}%
\label{sec:neubin}}

Currently \textit{FLAT} can utilize two different \texttt{NeuBin}
modules:
\ \texttt{NeuBin\_F2C} and \texttt{NeuBin\_F3C}. These implement 
the Hamiltonian
$\hat{H}$ and the neutrino flavor state  $|\psi\rangle$ for the
2-flavor and 3-flavor mixing schemes, respectively.

The \texttt{NeuBin\_F2C} module defines two classes, \texttt{NeuBin}
and \texttt{NeuPot}.  A \texttt{NeuBin} object contains two
complex variables. These variables correspond to the two complex
components of 
the flavor wavefunction 
\begin{equation}
\psi\equiv\left(\begin{array}{c}
a_{\nu_e}\\a_{\nu_x}
\end{array}\right),
\label{eq:psi-fl2}
\end{equation}
where $a_{\nu_\alpha}\equiv\langle\nu_\alpha|\psi\rangle$ 
is the amplitude for a neutrino in state $|\psi\rangle$
to be in flavor state $|\nu_\alpha\rangle$, and where the flavor index
$\alpha = e,x$, with $x$ representing either the $\mu$ or $\tau$ neutrino or a
linear combination of these. 
Class \texttt{NeuPot}, however, does not render 
the corresponding flavor-basis representation of the Hamiltonian
$\hat{H}$ as the familiar Hermitian matrix in the Schr\"odinger
equation.
In the 2-flavor mixing case, for example, the Hamiltonian would be a
$2\times2$ complex array
\begin{equation}
H=\left(\begin{array}{cc}
\langle\nu_e|\hat{H}|\nu_e\rangle & \langle\nu_e|\hat{H}|\nu_x\rangle \\
\langle\nu_x|\hat{H}|\nu_e\rangle & \langle\nu_x|\hat{H}|\nu_x\rangle
\end{array}\right).
\end{equation}
Instead, \texttt{NeuPot} stores the Hamiltonian in a three-component
real vector 
$\bi{h}$ which is defined by
\begin{equation}
H = h_0 + \bi{h}\cdot\bsigma. 
\label{eq:H-com}
\end{equation}
Here the three Cartesian components of the vector $\bsigma$ corresponds
to the  Pauli matrices:
\begin{equation}
\sigma_1=\left(\begin{array}{cc}
0 & 1 \\ 1 & 0
\end{array}\right),\,
\sigma_2=\left(\begin{array}{cc}
0 & -\rmi \\ \rmi & 0
\end{array}\right),\,
\mathrm{and}\
\sigma_3=\left(\begin{array}{cc}
1 & 0 \\ 0 & -1
\end{array}\right).
\end{equation}
The trace of matrix $H$ [$h_0$ in Equation (\ref{eq:H-com})]
is of no physical significance and is ignorable. This is because $h_0$
generates 
only an overall common phase in the neutrino flavor amplitudes and,
therefore,
is not relevant for neutrino oscillations or neutrino flavor
transformation.

Equation (\ref{eq:schroedinger}) has an approximate
solution for a small parameter step $\dt$,
\begin{eqnarray}
\fl\psi(t+\dt)\simeq\exp(-\rmi H \dt)\psi(t)
\label{eq:sol}\\
\fl\phantom{\psi(t+\dt)}=\left(\begin{array}{cc}
\cos(h\dt)-\rmi \tilde{h}_3\sin(h\dt) & 
-(\rmi \tilde{h}_1+\tilde{h}_2)\sin(h\dt)\\
-(\rmi \tilde{h}_1-\tilde{h}_2)\sin(h\dt) & 
\cos(h\dt)+\rmi \tilde{h}_3\sin(h\dt)
\end{array}\right)\psi(t),
\label{eq:sol-2fl}
\end{eqnarray}
where
\begin{equation}
h \equiv |\bi{h}| = \sqrt{h_1^2+h_2^2+h_3^2}
\end{equation}
and
\begin{equation}
\tilde{h}_i \equiv \frac{h_i}{h}.
\end{equation}
With a specified Hamiltonian $\hat{H}$ (a \texttt{NeuPot} object),  
Equation (\ref{eq:sol-2fl}) is employed in module \texttt{NeuBin\_F2C}
to evolve a \texttt{NeuBin} object by a small step $\dt$.
The procedure implied by Equation (\ref{eq:sol}) preserves the unitarity of
$\psi$ and becomes
exact when $H$ is time independent.


For a particular momentum mode $\lambda$, the neutrino (antineutrino)
density operator $\hat\rho_\lambda$ ($\hat{\bar\rho}_\lambda$)
provides the necessary information to construct a neutrino-neutrino
forward scattering background potential $\hat{H}_{\nu\nu}$ 
for a given test neutrino. The density operator for neutrino 
mode $\lambda$ is, for example,
$\hat\rho_\lambda=|\psi_\lambda\rangle\langle\psi_\lambda|$.
Potential $\hat{H}_{\nu\nu}$ is essentially a 
summation of densities operators
over all neutrino and antineutrino modes,
weighted by the directional dependence of the weak current-current Hamiltonian
(see, e.g., \cite{Duan:2006an}).
The contents of the density matrices are  stored
in  \texttt{NeuPot} objects.
In the \texttt{NeuBin\_F2C} module, this is done
by, for example, writing density matrix $\rho$ as
\begin{equation}
\rho=\frac{1}{2}+\bi{s}\cdot\bsigma,
\end{equation}
where 
\begin{equation}
\bi{s}=\left(\begin{array}{c}
\mathrm{Re}(a_{\nu_e}^* a_{\nu_x})\\
\mathrm{Im}(a_{\nu_e}^* a_{\nu_x})\\
\case{1}{2}(|a_{\nu_e}|^2- |a_{\nu_x}|^2)
\end{array}\right).
\end{equation}

The approach adopted in \textit{FLAT} for handling the full $3\times 3$
neutrino flavor evolution problem is essentially similar to the method
outlined above for the $2\times 2$ case. It is, of course, necessarily
more complicated. Like the \texttt{NeuBin\_F2C} module, the
\texttt{NeuBin\_F3C} module defines
a \texttt{NeuBin} class implementing neutrino flavor wavefunction
\begin{equation}
\psi\equiv\left(\begin{array}{c}
a_{\nu_e}\\a_{\nu_\mu}\\a_{\nu_\tau}
\end{array}\right),
\end{equation}
in obvious analogy to Equation (\ref{eq:psi-fl2}).
The \texttt{NeuPot} class in the \texttt{NeuBin\_F3C} module defines
9 real variables: $u_{11}$, $u_{22}$, $u_{33}$, $u_{12}$, $u_{23}$,
$u_{31}$, $w_{12}$, $w_{23}$, and $w_{31}$. Variables $u_{ij}$ and $w_{ij}$ are
defined in
\begin{equation}
H_{ij}=u_{ij}+\rmi w_{ij},
\end{equation}
where $H_{ij}$ are the elements of the $3\times3$ flavor-basis
Hamiltonian matrix
\begin{equation}
H=\left(\begin{array}{ccc}
\langle\nu_e|\hat{H}|\nu_e\rangle & \langle\nu_e|\hat{H}|\nu_\mu\rangle 
& \langle\nu_e|\hat{H}|\nu_\tau\rangle\\
\langle\nu_\mu|\hat{H}|\nu_e\rangle &
\langle\nu_\mu|\hat{H}|\nu_\mu\rangle 
& \langle\nu_\mu|\hat{H}|\nu_\tau\rangle\\
\langle\nu_\tau|\hat{H}|\nu_e\rangle &
\langle\nu_\tau|\hat{H}|\nu_\mu\rangle 
& \langle\nu_\tau|\hat{H}|\nu_\tau\rangle
\end{array}\right).
\end{equation}

The $3\times 3$ module \texttt{NeuBin\_F3C} also uses 
Equation (\ref{eq:sol}) to evolve 
a neutrino flavor wavefunction over a short step $\dt$.
However, it is difficult to express $\exp(-\rmi H\dt)$ for the
$3\times 3$ case in an explicit form like Equation (\ref{eq:sol-2fl}).
Instead, the module \texttt{NeuBin\_F3C} solves
the eigenvalue equations
\begin{equation}
\sum_j H_{ij} V_{jk} = \zeta_k V_{ik},\quad i,j,k=1,2,3,
\label{eq:eigen}
\end{equation}
and uses
\begin{equation}
\exp(-\rmi H\dt)=V\left(\begin{array}{ccc}
\rme^{-\rmi\zeta_1\dt} & & \\
& \rme^{-\rmi\zeta_2\dt} & \\
& & \rme^{-\rmi\zeta_3\dt}
\end{array}\right)V^\dagger.
\end{equation}
Here $V_{ik}$ represents the $i$th component of the $k$th eigenvector of the
Hamiltonian $H$, with corresponding  eigenvalue $\zeta_k$.

Module \texttt{NeuBin\_F3C} employs an analytical
method to solve Equation (\ref{eq:eigen}). This method
is based on the algorithm developed
in \cite{Kopp:2006wp} and takes into account the fact
that the trace term in $H$ is not relevant for neutrino flavor
evolution. Assuming that 
\begin{equation}
u_{11}+u_{22}+u_{33}=0
\end{equation}
and, if necessary, shifting $u_{ii}$ to make this sum 0, we define
\begin{eqnarray}
\fl p=-3[(u_{22}u_{33} - u_{11}^2) - (u_{12}^2+w_{12}^2)
-(u_{23}^2+w_{23}^2) -(u_{31}^2+w_{31}^2)],\\
\fl q=-\case{27}{2}[u_{11}(u_{23}^2+w_{23}^2)
+u_{22}(u_{31}^2+w_{31}^2) +u_{33}(u_{12}^2+w_{12}^2)
-u_{11}u_{22}u_{33}\nonumber\\
\fl\phantom{q=}
+2(-u_{12}u_{23}u_{31}+u_{12}w_{23}w_{31}+w_{12}u_{23}w_{31}+w_{12}w_{23
}u_{31})],\\
\fl\phi=\arccos(q).
\end{eqnarray}
The eigenvalues can be written as
\numparts
\begin{eqnarray}
\zeta_1&=&\frac{2\sqrt{p}}{3}\cos\phi,\\
\zeta_2&=&\frac{2\sqrt{p}}{3}\cos\left(\phi+\frac{2\pi}{3}\right),\\
\zeta_3&=&-\zeta_1-\zeta_2.
\end{eqnarray}
\endnumparts

From Equation (\ref{eq:eigen}) it can be seen that the eigenvector $\bi{V}_i$
(with corresponding eigenvalue $\zeta_i$) is orthogonal to the complex
conjugates of the row vectors
in matrix $H-\zeta_i I$. Assuming that $\bi{X}^{(1)}$ and $\bi{X}^{(2)}$
are two linearly independent row vectors in matrix $H-\zeta_i I$, we can
write
\begin{equation}
\bi{V}_i=\frac{\bi{X}^{(1)}\times\bi{X}^{(2)}}%
{|\bi{X}^{(1)}\times\bi{X}^{(2)}|}.
\end{equation}
Because the eigenvectors of $H$ are orthogonal to each other,
it is necessary to solve for only two eigenvectors, say $\bi{V}_1$ and
$\bi{V}_2$.
The third eigenvector, $\bi{V}_3$, can be determined from
\begin{equation}
\bi{V}_3=\bi{V}_1^*\times\bi{V}_2^*.
\end{equation}

\subsection{\texttt{NBGroup\_PM} modules and the parallelization of
\textit{FLAT}%
\label{sec:nbgroupmm}}

Module \texttt{NBGroup\_MM} is a single-thread version of an
\texttt{NBGroup\_PM} module. This single thread implementation uses an
algorithm, similar to
the modified midpoint method \cite{Press:2002oc}, to determine
$|\psi_\lambda(r+\dr)\rangle$ from $|\psi_\lambda^{(0)}(r)\rangle$ 
according to Eq.~(\ref{eq:schroedinger}). 
(Note that the relation between the radial coordinate $r$
and the Affine parameter value $t$ for each mode $\lambda$
depends on the trajectory direction of this neutrino mode.) 
The algorithm consists of three parts.
In the first part, an approximation to $|\psi_\lambda(r+\dr)\rangle$
is obtained using radial coordinate step size $\dr$:
\newcounter{AgStep}
\newcounter{AgLastStep}
\begin{list}{{Step \arabic{AgStep}:}\,}{\usecounter{AgStep}}
\item \label{ag:p1-sumi}Call \texttt{NBGroup::set\_bg()}
to set the background Hamiltonian
$\hat{H}_\bg^{(0)}(\vartheta_0)\equiv\hat{H}_\matt(n_e(r))
+\hat{H}_{\nu\nu}(\psi^{(0)},r,\vartheta_0)$ 
for each neutrino trajectory $\vartheta_0$.
This function utilizes flavor states $|\psi_{\lambda^\prime}^{(0)}(r)\rangle$
for all neutrino modes $\lambda^\prime$ in the problem.

\item \label{ag:p1-evi}Find 
$|\psi_\lambda^{(1)}(r+\dr)\rangle:=
\hat{U}_\lambda(\hat{H}_\bg^{(0)},r,\dr)
|\psi_\lambda^{(0)}(r)\rangle$ for each neutrino mode $\lambda$.
Here $\hat{U}_\lambda(\hat{H}_\bg^{(0)},r,\dr)$ 
is the operator
evolving $|\psi_\lambda\rangle$ through a small step. 
This ``parameter evolution operator'' is implemented as
\texttt{NeuBin::evolve()}. 
The step in mode $\lambda$'s Affine parameter $t_{\vartheta_0(\lambda)}$ 
corresponding to $\dr$ is 
$\dt_{\vartheta_0}(r,\dr)=t_{\vartheta_0}(r+\dr)-t_{\vartheta_0}(r)$.

\item \label{ag:p1-sumf}Compute
$\hat{H}_\bg^{(1)}(\vartheta_0)\equiv\hat{H}_\matt(n_e(r+\dr))
+\hat{H}_{\nu\nu}^{\vartheta_0}(\psi^{(1)},r+\dr,\vartheta_0)$ 
for each neutrino trajectory $\vartheta_0$.

\item \label{ag:p1-evf}Find
$|\psi_\lambda^{(2)}(r+\dr)\rangle:=
\hat{U}_\lambda(\hat{H}_\bg^{(1)},r,\dr)
|\psi_\lambda^{(0)}(r)\rangle$ for each neutrino mode $\lambda$.

\item \label{ag:p1-avg}Find $|\psi_\lambda^{(3)}(r+\dr)\rangle$ 
as an average of 
$|\psi_\lambda^{(1)}(r+\dr)\rangle$ and
$|\psi_\lambda^{(2)}(r+\dr)\rangle$ for each neutrino mode $\lambda$.

\setcounter{AgLastStep}{\value{AgStep}}
\end{list}

The second part of the algorithm essentially repeats the first part,
but with the step size reduced to $\case{1}{2}\dr$:
\begin{list}{{Step \arabic{AgStep}:}\,}{\usecounter{AgStep}}
\setcounter{AgStep}{\value{AgLastStep}}
\item \label{ag:p2-evi}Find 
$|\psi_\lambda^{(4)}(r+\case{1}{2}\dr)\rangle:=
\hat{U}(\hat{H}_\bg^{(0)},r,\case{1}{2}\dr)
|\psi_\lambda^{(0)}(r)\rangle$.

\item \label{ag:p2-summ}Compute 
$\hat{H}_\bg^{(4)}(\vartheta_0)\equiv\hat{H}_\matt(n_e(r+\case{1}{2}\dr))
+\hat{H}_{\nu\nu}(\psi^{(4)},r+\case{1}{2}\dr,\vartheta_0)$ 
for each neutrino trajectory $\vartheta_0$.

\item \label{ag:p2-evm}Find 
$|\psi_\lambda^{(5)}(r+\dr)\rangle:=
\hat{U}(\hat{H}_\bg^{(4)},r,r+\dr)
|\psi_\lambda^{(0)}(r)\rangle$.

\item \label{ag:p2-sumf}Compute 
$\hat{H}_\bg^{(5)}(\vartheta_0)\equiv\hat{H}_\matt(n_e(r+\dr))
+\hat{H}_{\nu\nu}^{\vartheta_0}(\psi^{(5)},r+\dr,\vartheta_0)$ 
for each neutrino trajectory $\vartheta_0$.

\item \label{ag:p2-evf}Find 
$|\psi_\lambda^{(6)}(r+\dr)\rangle:=
\hat{U}(\hat{H}_\bg^{(5)},r+\case{1}{2}\dr,\case{1}{2}\dr)
|\psi_\lambda^{(4)}(r+\case{1}{2}\dr)\rangle$.

\item \label{ag:p2-avg}Find $|\psi_\lambda^{(7)}(r+\dr)\rangle$ as an
average of $|\psi_\lambda^{(5)}(r+\dr)\rangle$ and
$|\psi_\lambda^{(6)}(r+\dr)\rangle$ for each neutrino mode $\lambda$.
\setcounter{AgLastStep}{\value{AgStep}}
\end{list}

The last part of the algorithm estimates the numerical error
$\epsilon$ and determines the course of subsequent computation:
\begin{list}{{Step \arabic{AgStep}:}\,}{\usecounter{AgStep}}
\setcounter{AgStep}{\value{AgLastStep}}
\item \label{ag:err}Estimate the numerical error $\epsilon$ to be
$\case{1}{3}$
of the maximum of the differences between 
$|\psi_\lambda^{(3)}(r+\dr)\rangle$ and
$|\psi_\lambda^{(7)}(r+\dr)\rangle$ 
in all neutrino modes.
\item \label{ag:stp}Set the new step size to be 
$\dr:=\xi \sqrt{\epsilon_0/\epsilon}$,
where $\xi$ is an empirical constant, and
$\epsilon_0$ is the prescribed error tolerance.
\item \label{ag:dec}Set 
$|\psi_\lambda(r+\dr)\rangle:=|\psi_\lambda^{(7)}(r+\dr)\rangle$ if 
$\epsilon\leq\epsilon_0$. 
Return to Step \ref{ag:p1-sumi} and repeat the algorithm as necessary.
\end{list}

Module \texttt{NBGroup\_MM\_SMP} is a parallelized version of
\texttt{NBGroup\_MM} using Pthreads.
In \texttt{NBGroup\_MM\_SMP}, the algorithm described above
is rearranged into a multi-stage pipeline through which all neutrino
states must flow. Each working thread first
enters a thread-mutual-exclusive routine
\texttt{NBGroup\_PM::new\_task()}.
In this routine, the thread
is assigned a subset of neutrino states $|\psi_\lambda\rangle$ at
a particular stage. After finishing the operations on
$|\psi_\lambda\rangle$
at this stage, the thread re-enters  
\texttt{NBGroup\_PM::new\_task()}. Subsequently, the thread will
be reassigned with another subset consisting of any remaining neutrino
states at the same stage.
If there are no remaining states, and if the dependencies are clear, 
\texttt{NBGroup\_PM::new\_task()} will assign to the working thread a subset of
neutrino states at the next stage in the calculation. 
The stages are arranged so that there is no dependence between
neighboring stages except the last two. This is done as follows:
\begin{list}{{Stage \arabic{AgStep}:}\,}{\usecounter{AgStep}}
\item Step \ref{ag:p2-evi} in the above algorithm.
The last working thread at this stage completes step \ref{ag:p2-summ}.
\item Step \ref{ag:p1-evi}. The last thread also completes step
\ref{ag:p1-sumf}.
\item Step \ref{ag:p2-evm}. The last thread also completes step
\ref{ag:p2-sumf}.
\item Stpdf \ref{ag:p1-evf} and \ref{ag:p1-avg}.
\item Stpdf \ref{ag:p2-evf} and \ref{ag:p2-avg}.
\item Stpdf \ref{ag:err}, \ref{ag:stp}, \ref{ag:dec} and
\ref{ag:p1-sumi}.
\end{list}

\textit{FLAT} has been designed to take
advantage of computing nodes which are capable of running multiple
threads simultaneously, e.g., those nodes with multi-cores,
multiple-CPU's and/or
Simultaneous MultiThreading (SMT) technology. Multiple threads
can be exploited by \textit{FLAT} by
swapping out the \texttt{NBGroup\_MM} module
and replacing it with  \texttt{NBGroup\_MM\_SMP}. Because Pthreads share
a common memory space and do not need to pass messages to each other, 
the parallelization of \textit{FLAT} using \texttt{NBGroup\_MM\_SMP}
is quite efficient. On an IBM 32-way Power4 computing node,  
multi-angle code with 32 working threads can run as
fast as $\sim31.2\times$ the speed of the single-threaded version. 
On a single-CPU Pentium4 desktop computer 
with Hyper-Threading technology (HT) enabled, threaded single-angle 
calculations can run $\gtrsim30\%$ faster than non-threaded
calculations.

In order to utilize multiple computing nodes, \textit{FLAT} is also
parallelized using Message Passing Interface (MPI) \cite{MPI}
in a manner
similar to  the implementation in \textit{BULB} (see Section \ref{sec:bulb}).
This has been done by rewriting portions of the \texttt{NBGroup\_EM}
and \texttt{NBGroup\_IO} modules. For example,  function
\texttt{NBGroup\_EM::set\_bg()} of 
the \texttt{NBGroup\_PNS2D} module can be modified
to compute the neutrino self-coupling $\hat{H}_{\nu\nu}$ based on
not only the neutrino states on the residing node but also
the states on other nodes. Because \texttt{NBGroup\_PM} modules
are untouched in this scheme, an MPI/Pthread hybrid parallel model can
be realized
by choosing the \texttt{NBGroup\_MM\_SMP} module together with the
MPI version of the \texttt{NBGroup\_PNS2D} and 
\texttt{NBGroup\_NETCDF} modules.

\subsection{The structure of \textit{BULB}%
\label{sec:bulb}}

   We have developed a second code called \textit{BULB}
to independently
verify the results of the first code.  \textit{BULB}, developed at LANL, 
is written in 
Fortran90 and uses the MPI library to communicate between processes.
As such it is suitable for use on many modern large-scale homogeneous computers.

\textit{BULB} is parallelized by splitting neutrino trajectories into
different groups of emission angles.
In this implementation, each node independently handles all neutrino energy
variables for each angle group.
This is acceptable for machines of up to 500--1000 nodes, as the largest
calculations can be split into one angle per processor.  For larger 
machines it would also be desirable to split different energy bins among
different processors.  This may be needed in the future for more complex,
non-spherical
geometries where more than roughly 500 energy bins and 500 angle
bins may be required.

   The numerical integration algorithm on each node is a fairly simple
treatment of nonlinear coupled differential equations.  For each step,
we choose an initial step size $\Delta r$, evaluating the matter background
potential $H_\matt$ as an average over the initial and final point.  
In analogy to the procedure described above for \textit{FLAT},
this increment in radius can be translated into the increment
in the Affine parameter $\Delta t$ along each individual trajectory.

Initially, we assume that the background neutrino potential is constant 
over the step.  The overall Hamiltonian $H$ 
can then be diagonalized efficiently 
for $2\times2$ or $3\times3$ neutrino cases, and 
the propagated wave function at
the end of the step is 
$\psi_\mathrm{fin}=\exp(-\rmi H \Delta t)\psi_\mathrm{ini}$
(see Section \ref{sec:neubin}). 
When the neutrino-neutrino scattering contribution is small,
this formulation can allow for significant propagation distances.
The step sizes in this case are
bounded by the scale height of the matter profile
or the length scales related to
the fluctuations in the background electron potential.
This formulation also preserves the unitarity of the neutrino density matrix.

Upon completion of this initial propagation, the final wave function
is used to compute a revised background neutrino potential 
$H_{\nu\nu}$ at the final point.
This new potential is then averaged with the initial potential to 
produce a neutrino potential averaged over the propagation step, similar to the
treatment of the electron background potential. An improved
final wave function is obtained by propagating with this averaged
interaction.  In principle this averaging scheme can be iterated, though in
practice it is nearly always converged after the first iteration.

As in \textit{FLAT}, in \textit{BULB}
we check for convergence with respect to the step size by comparing
the final density matrix obtained in a single step $\delta r$ with
that obtained by taking two stpdf with half the step size. If the
density matrix does not match within a predefined tolerance, the
step size is reduced and the process is repeated.  Periodically,
we attempt to increase the step size in order to efficiently treat
the large radius regions where the neutrino background potential
and the electron density are fairly small.

A clear limitation in all these problems is that for each step
one must calculate the full neutrino density matrix and the
angular dependence of the neutrino potential for each 
energy and angle. This involves a sum over all energies,
which in the present version of \textit{BULB} is performed independently on each node.
Calculating neutrino potentials 
also involves a sum over angles, as the neutrino interaction
depends upon $\cos \vartheta_\bi{pq}$, where $\vartheta_\bi{pq}$ 
is the angle between two neutrinos with momenta $\bi{p}$
and $\bi{q}$, respectively.

For spherical symmetry the angular dependence can be retained by
evaluating the following sums on each node:
\numparts
\begin{eqnarray}
\bar{\rho}_n  & = & \sum_{j=1}^{N_\mathrm{th}} \sum_{i=1}^{N_\mathrm{e}} 
\rho(i,j) \\
\bar{\rho}^\prime_n & = & \sum_{j=1}^{N_\mathrm{th}} 
\cos{\vartheta_j} \sum_{i=1}^{N_\mathrm{e}} \rho(i,j),
\end{eqnarray}
\endnumparts
where index $n$ labels the node, $\vartheta_j$ is the angle between
the $j$'th neutrino trajectory and the radial direction at the current radius,
and the sum $i$ and $j$ run over the number of energy grid points $N_\mathrm{e}$
and the number of angles on the local node $N_\mathrm{th}$, respectively.  
 The full neutrino potential for all angles can then be evaluated by
using MPI global sum calls to sum both the angle independent terms $\bar\rho$
and the  angle-dependent terms $\bar\rho^\prime$ and 
by using the formula
\begin{equation}
\cos\vartheta_{ij} = \cos \vartheta_i \cos \vartheta_j + 
\sin\vartheta_i \sin\vartheta_j \cos (\phi_i - \phi_j),
\end{equation}
where $\phi_i$ and $\phi_j$ are the azimuthal angles
 of the two beams. In spherical symmetry
the second term vanishes because of the averaging over $(\phi_i - \phi_j)$
\cite{Duan:2006an}.

Convergence of the overall calculation is carefully checked by comparing
results with different error tolerances and their associated step sizes,
and by comparing results with different numbers of energy and angle bins.
Although qualitative results can often be obtained with a modest
number of angular bins, a rather large number of energy and angle
bins are required to get stable solutions.  This can typically involve
several hundred thousand coupled non-linear differential equations.

A ``typical'' calculation such as those described in the results section below
involve $\sim(800\,\mathrm{angle\, bins})\times(400\,\mathrm{energy\, bins})$.  
Near the surface of the proto-neutron
star, the wavelengths are very small and integration step sizes can be
as small as one millimeter.  By the time one reaches 20 km, or about twice
the radius of the star, the relevant wavelengths have increased and the
step size is of order one centimeter. The total computational time required
for a single calculation is approximately 8000 CPU hours on a typical
cluster with 2 GHz AMD Opteron processors. The computational time 
is roughly equally
divided between computing the evolution on each node 
and gathering the neutrino
densities across all nodes.  Storage of flavor states for
1000 neutrino modes
at various radii for later analysis requires approximately 
4 GB of storage space.

For understanding the propagation of the neutrinos from small to large
distances, \textit{BULB} can also dump out the neutrino potentials or
the full wave function at user-defined spatial intervals.  Storing
the full wave function is rather slow, so this is done less often than
sums over the wave function or density matrix.  The latter are sufficient
to determine the rough survival probabilities that could be detected in
a terrestrial observation.

\section{Results%
\label{sec:results}}

The first simulations carried out with both codes
focused on the late time, hot bubble epoch of the core collapse
supernova phenomenon. In this post-explosion environment,
the matter density surrounding the proto-neutron star is high, but
still lower than in the earlier shock re-heating epoch.
In the hot bubble, the neutrino self-coupling Hamiltonian
$\hat{H}_{\nu\nu}$
can be dominant in some regions and sub-dominant 
relative to the matter potential in others.

In our calculations
we can adopt a variety of matter density profiles,
neutrino and antineutrino luminosities, and
neutrino and antineutrino energy spectra.
For the specific example problem of late-time supernova
neutrino flavor evolution, we adopt a simple analytical profile
for electron number density:
\begin{equation}
\fl n_e\simeq 
(1.6\times10^{36}\,\mathrm{cm}^{-3})Y_e
\exp\left(\frac{R_\nu-r}{0.18\,\mathrm{km}}\right)
+(6.0\times10^{30}\,\mathrm{cm}^{-3})Y_e
\left(\frac{10\,\mathrm{km}}{r}\right)^3,
\end{equation}
where electron fraction is $Y_e=0.4$ and the
radius of the neutrino sphere is $R_\nu=11$ km.
We assume that initial energy spectra for neutrinos
are of two-parameter black body form
\begin{equation}
f_\nu(E_\nu)=\frac{1}{F_2(\eta_\nu)}\frac{1}{T_\nu^3}
\frac{E_\nu^2}{\exp(E_\nu/T_\nu-\eta_\nu)+1},
\end{equation}
where $\eta_\nu$ is the degeneracy parameter,
$T_\nu$ is the neutrino temperature, and
\begin{equation}
F_k(\eta)\equiv\int_0^\infty
\frac{x^k\rmd x}{\exp(x-\eta)+1}.
\end{equation}
For the late-time supernova environment
we take $\langle E_{\nu_e}\rangle=11$ MeV,
$\langle E_{\bar\nu_e}\rangle=16$ MeV,
$\langle E_{\nu_\tau}\rangle=\langle E_{\bar\nu_\tau}\rangle=25$ MeV,
and $\eta_{\nu_e}=\eta_{\bar\nu_e}=\eta_{\nu_\tau}=\eta_{\bar\nu_\tau}=3$.
Here by $\nu_\tau$ and $\bar\nu_\tau$ we actually mean
linear combinations of neutrinos and antineutrinos of
the real $\mu$ and $\tau$ flavors, respectively. 
With these choices we have
$T_{\nu_e}\simeq2.76$ MeV, $T_{\bar\nu_e}\simeq4.01$ MeV, 
and $T_{\nu_\tau}=T_{\bar\nu_\tau}\simeq6.26$ MeV.

For our example $2\times2$ mixing treatment presented below,
we have taken the effective mixing angle to be
$\theta=0.1$, and mass-squared difference to be
$\Delta m^2=\pm3\times10^{-3}\,\mathrm{eV}^2$.
Here the plus sign defines a so-called ``normal neutrino mass hierarchy'',
and the minus sign defines an ``inverted neutrino mass hierarchy''.
There are six parameters in the full $3\times3$ vacuum mixing problem.
These parameters consist of two mass-squared differences
($\Delta m_{12}^2$ and $\Delta m_{13}^2$),
three mixing angles ($\theta_{12}$, $\theta_{13}$ and
$\theta_{23}$) and the \textit{CP}-violating phase $\delta$.
Mass-squared differences $\Delta m_{12}^2$ and $|\Delta m_{13}^2|$
and mixing angles $\theta_{12}$ and $\theta_{23}$ have been measured
outright, while $\delta$ and the neutrino mass hierarchy 
are as yet unmeasured.
In our $2\times2$ mixing treatment 
$\Delta m^2\simeq \Delta m_{13}^2$ and $\theta\simeq\theta_{13}$.
Although $\theta_{13}$ also remains unmeasured, current laboratory results
suggest that $\sin^2(2\theta_{13})\lesssim0.1$.

Our preliminary single-angle 2-flavor simulations
gave very puzzling results. Contrary to the studies based on
the MSW effect (e.g., 
\cite{Kuo:1987qu,Dighe:1999bi,Lunardini:2003eh,Kneller:2005hf}), 
these calculations showed that the flavor evolution histories of
neutrinos with
different energies could be coupled together and that all neutrinos and
antineutrinos
could experience collective flavor transformation. This could be true
in our calculations even when the neutrino-neutrino forward scattering
potential 
was sub-dominant.

Analysis showed that the collective flavor transformation observed in
our simulations was
not the well known synchronized flavor transformation seen in earlier
single-angle calculations (e.g.,
\cite{Pastor:2002we,Balantekin:2004ug}). In synchronized flavor
transformation,
neutrinos and antineutrinos of all energies can undergo a common
MSW-like
flavor transformation in the region where a neutrino with a
representative energy would experience a conventional MSW resonance.
In addition, for neutrino-antineutrino gases that are dominated by neutrinos,
synchronized flavor transformation can occur only in the
normal neutrino mass hierarchy case. In our single-angle calculations,
however, collective flavor transformation occurred for both the normal and
inverted neutrino mass hierarchies and exhibited oscillatory behaviors
over a relatively wide radius range (see, e.g., Figure 8
in \cite{Duan:2006an}). 

Based on these preliminary results, we re-investigated 
collective flavor transformation in dense neutrino-antineutrino gases
in the presence of ordinary matter \cite{Duan:2005cp}.
We found that the oscillatory collective flavor transformation
could be explained as ``bipolar oscillations''. This
 type of collective flavor transformation was first studied in the 
context of the
early universe \cite{Kostelecky:1993dm,Kostelecky:1994dt,Samuel:1995ri}.

Contrary to earlier widely-accepted conclusions, our calculations and 
analyses showed that collective
flavor transformation of the bipolar type can appear even in
the presence of a large and dominant matter density. This is a
surprising and 
paradigm-changing result. In the past the astrophysics community
had assumed that neutrino flavor transformation 
in the high matter density environment near the 
neutron star would be suppressed (e.g., \cite{Pastor:2002we}).
However, our results now compelled us
 to conclude that large scale conversion of 
neutrino flavors could take place in such an environment. 
This conclusion was particularly surprising given the 
fact that the experimentally-inferred neutrino mass-squared 
differences are small.

The neutrino and antineutrino trajectories in our calculations are
labeled by emission angle $\vartheta_0$. This is the angle 
between the trajectory and the normal 
to the neutrino sphere surface at the neutrino's or antineutrino's point
of origin on this surface. For example, $\vartheta_0 = 0$ 
corresponds to a radially-directed neutrino trajectory, 
while $\vartheta_0=\pi/2$ corresponds to a neutrino 
moving tangentially to the neutrino sphere.

In the single-angle approximation, neutrinos and antineutrinos
are assumed to have the
same flavor evolution histories as those with the same energies but
traveling along radial trajectories. To test the validity of 
this approximation, we carried out
multi-angle calculations under the same conditions
used in the single-angle cases, except of course for
the obvious geometric issues associated with binning the trajectory
directions \cite{Duan:2006an,Duan:2006jv}. 

We first consider the artificial case where we ignore neutrino
self-coupling and consider only matter-driven neutrino and antineutrino flavor
evolution. This is achieved by setting neutrino luminosities
$L_{\nu_e}=L_{\bar\nu_e}=L_{\nu_\tau}=L_{\bar\nu_\tau}=0$ erg/s.
This case corresponds to
 the  ``MSW'' case which had been the 
paradigm for many if not most studies of neutrino flavor transformation
effects for the supernova neutrino signal 
(e.g., \cite{Kuo:1987qu,Dighe:1999bi,Lunardini:2003eh,Kneller:2005hf})
and for nucleosynthesis 
(e.g., \cite{Heger:2003mm,Yoshida:2006sk,Yoshida:2006qz}).
As we will see,
this case fails to capture the essential physics of the problem.

\begin{figure}[t]
\href{http://simulations4snu.googlepages.com/late_time_2flavor_nh_L0.html}{%
\includegraphics*[origin=bl,width=\textwidth,keepaspectratio]{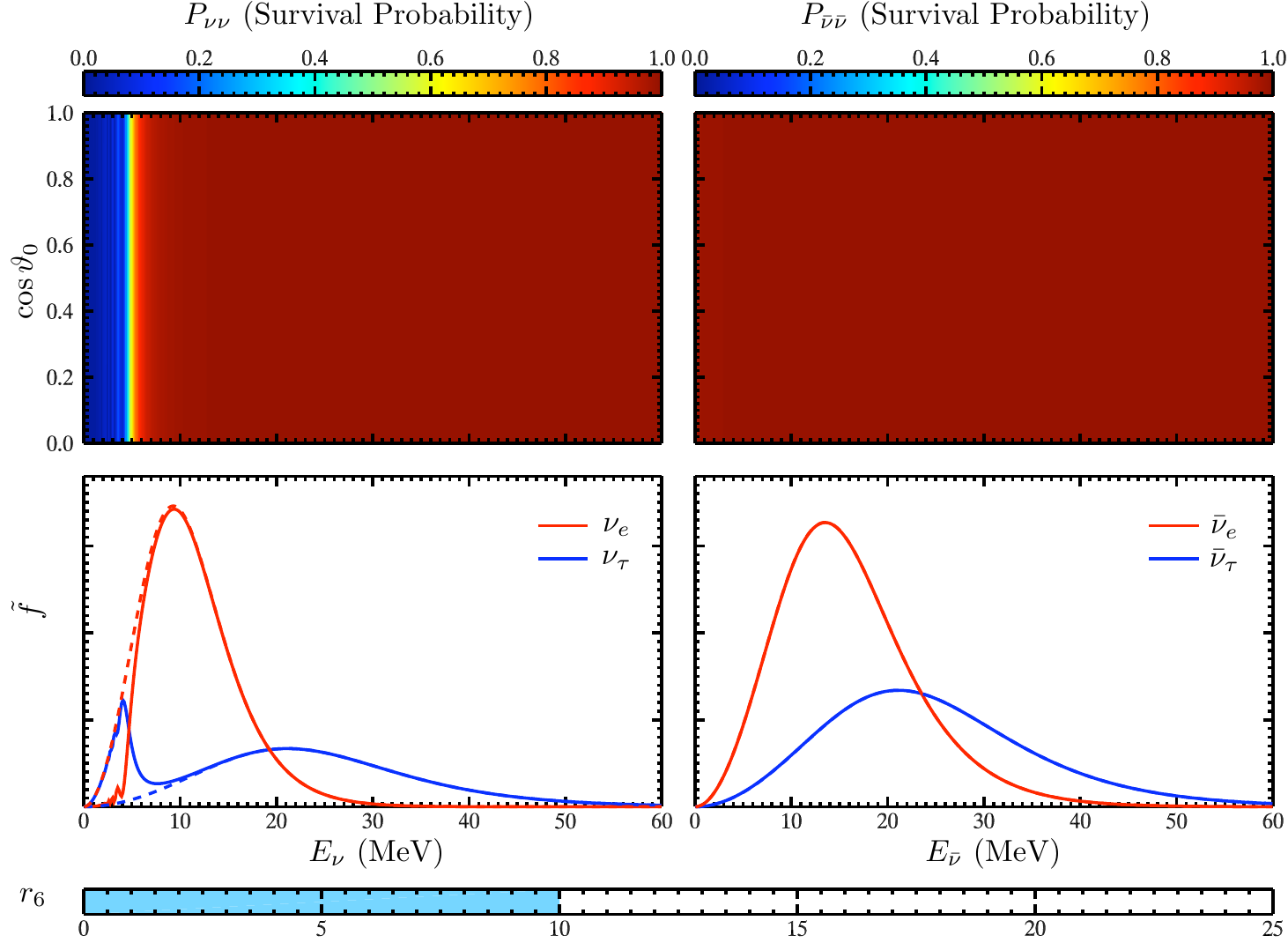}}%
\caption{\label{fig:normalMSW}%
For the pure matter-driven MSW case with the 
normal neutrino mass hierarchy we show
neutrino survival probability $P_{\nu\nu}$  (upper left panel)
and antineutrino survival probability $P_{\bar\nu\bar\nu}$  (upper right panel)
as functions of neutrino energy $E_\nu$
or antineutrino energy $E_{\bar\nu}$ and cosine of the emission angle, 
$\cos\vartheta_0$,
for values of $r_6$ (radius in units of $10^6\,\mathrm{cm}$). 
The progress bar at the bottom shows $r_6$.
Survival probability is given by the color code
at the top of the panels; red corresponding to little or 
no flavor transformation,
and blue corresponding to complete flavor conversion.
Note that $\cos\vartheta_0$ is unity for radially-propagating 
particles, and zero
for tangentially-propagating ones. 
In the bottom panels, the corresponding
angle-averaged energy distribution functions $\tilde{f}$
(arbitrary normalization) are shown
for neutrinos (lower left panel) or antineutrinos (lower right panel) 
for flavor  $e$ ($\tau$) by the red (blue) curves.
The dashed and solid curves in the bottom panels
are for the neutrino energy spectra at the neutrino sphere
and at the radius indicated by the progress bar, respectively. 
This figure corresponds
to the snapshot at $r_6\simeq10$ in the 
\href{http://simulations4snu.googlepages.com/late_time_2flavor_nh_L0.html}%
{full simulation}.}
\end{figure}

The matter density is assumed here to be spherically symmetric. As a
result,
we expect the pure matter-driven MSW case to exhibit little dependence
on neutrino or antineutrino
emission angle (trajectory direction). 
Figure \ref{fig:normalMSW} shows a snapshot ($r\simeq100$ km) 
of one of our simulations for this case.
A \href{http://simulations4snu.googlepages.com/late_time_2flavor_nh_L0.html}%
{movie} presenting the full simulation is available for downloading.
There are several prominent 
features of neutrino and antineutrino flavor transformation in this
scenario. We see no transformation of antineutrinos in this case, and
neutrinos are transformed only at relatively low energy $E_\nu$. This is
classic MSW behavior: because the neutrino mass-squared difference is
small, we must go to large radius before neutrinos with average energies
are affected, and antineutrinos never are.
 And, indeed, there is little
emission angle-dependence in this case.

\begin{figure}[t]
\href{http://simulations4snu.googlepages.com/late_time_2flavor_ih_L0.html}{%
\includegraphics*[origin=bl,width=\textwidth,keepaspectratio]%
{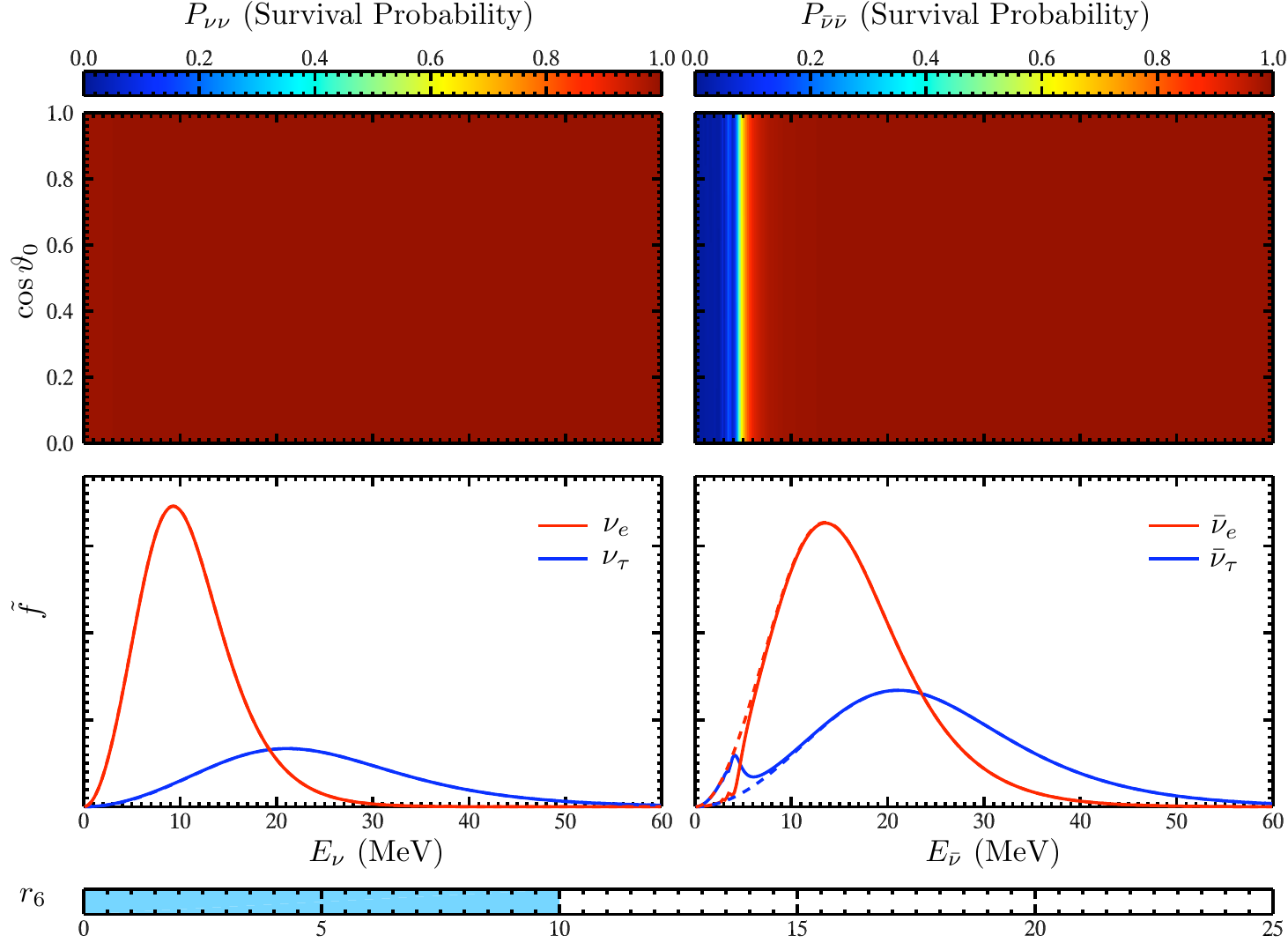}}
\caption{\label{fig:invertedMSW}%
This figure is the same as Figure \ref{fig:normalMSW},
except now we show neutrino and antineutrino flavor transformation
results 
for the pure matter-driven MSW case with the 
inverted neutrino mass hierarchy. 
This figure corresponds to the snapshot at $r_6\simeq10$ in the 
\href{http://simulations4snu.googlepages.com/late_time_2flavor_ih_L0.html}{
full simulation}.}
\end{figure}

Figure \ref{fig:invertedMSW} shows a snapshot ($r\simeq100$ km) 
of one of our simulations for the pure
matter-driven MSW case,
but now for the inverted neutrino mass hierarchy. 
A \href{http://simulations4snu.googlepages.com/late_time_2flavor_ih_L0.html}%
{movie} presenting the full simulation is available for downloading.
These results are in
some sense similar to those for the normal neutrino mass hierarchy case,
except that here it is the low energy {\it antineutrino} flavors that
are transformed. There is no neutrino flavor conversion in the inverted
neutrino mass hierarchy case.

We can repeat these calculations for the same supernova model,
but now with the full neutrino interaction Hamiltonian,
including both the matter potential and the neutrino self-coupling
potential $\hat{H}_{\nu\nu}$. The resulting neutrino and antineutrino
flavor evolution
is dramatically different from that in the MSW cases.

\begin{figure}[t]
\href{http://simulations4snu.googlepages.com/late_time_2flavor_nh.html}{%
\includegraphics*[origin=bl,width=\textwidth,keepaspectratio]%
{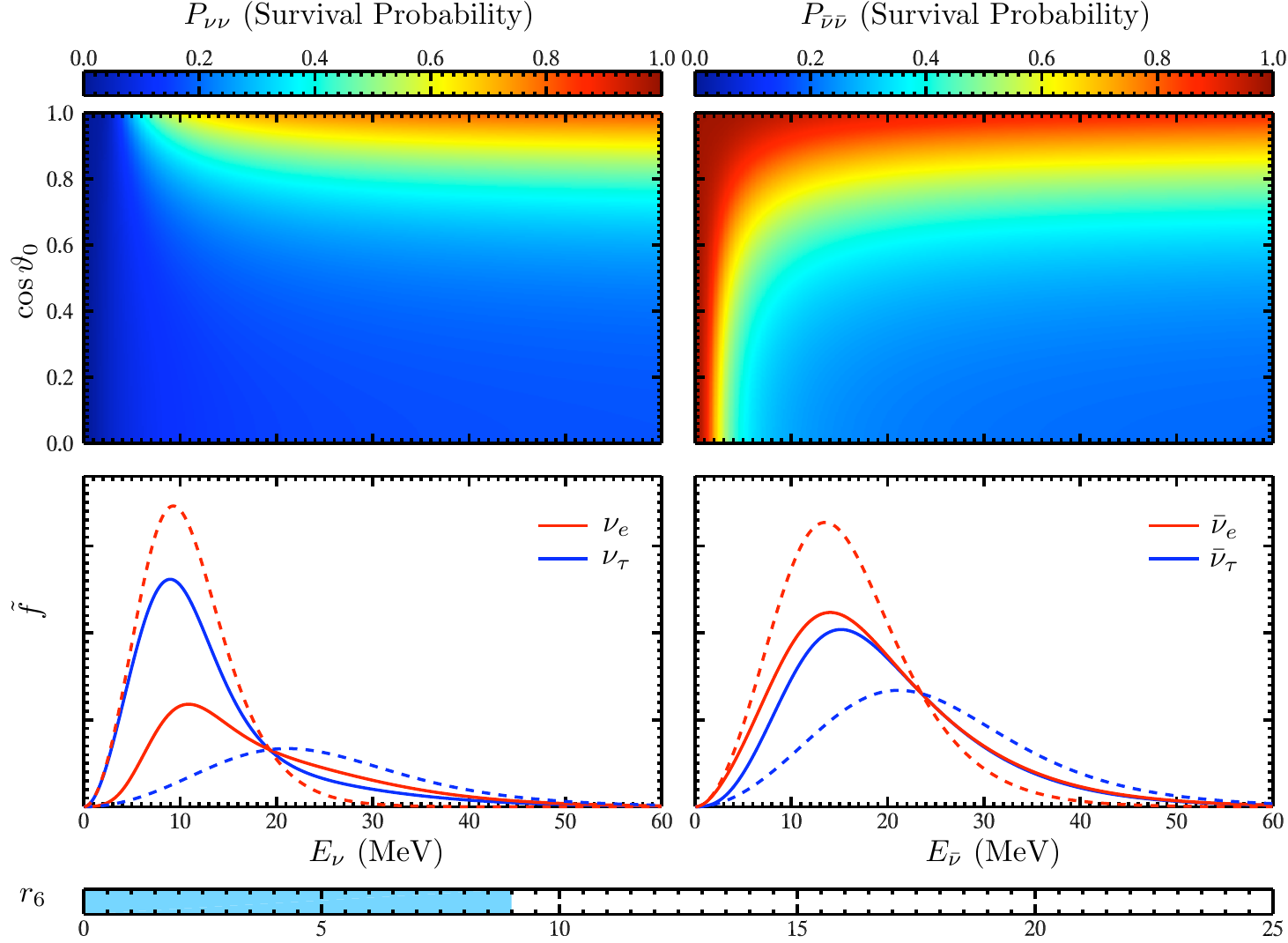}}
\caption{\label{fig:normal}%
Neutrino and antineutrino flavor transformation results 
for the normal neutrino mass hierarchy case
but now with full neutrino self-coupling 
and trajectory coupling.
Setups and definitions are the same as in Figure \ref{fig:normalMSW}.
Here we take the luminosities of all neutrino species 
 to  be $10^{51}$ erg/s.
This figure corresponds
to a snapshot at $r_6\simeq9.0$ in the 
\href{http://simulations4snu.googlepages.com/late_time_2flavor_nh.html}{%
full simulation}.}
\end{figure}

Figure \ref{fig:normal} shows a snapshot ($r\simeq90$ km)
of one of our simulations for the
normal neutrino mass hierarchy case but now with
the full neutrino self-coupling Hamiltonian. 
A \href{http://simulations4snu.googlepages.com/late_time_2flavor_nh.html}%
{movie} presenting the full simulation is available for downloading.
In this simulation
we have taken all neutrino species to have equal luminosities and
$L_{\nu_e}=L_{\bar\nu_e}=L_{\nu_\tau}=L_{\bar\nu_\tau}=10^{51}$ erg/s. 
This simulation shows that at values of radius $r\sim90$ km
both neutrinos and
antineutrinos over broad energy ranges experience 
large-scale simultaneous flavor transformation.
Although earlier analytical studies \cite{Fuller:2005ae,Duan:2005cp}
 suggested that this behavior was possible,
this is nevertheless a surprising result as previous numerical simulations
did not find this behavior in comparable  or
earlier epochs \cite{Pastor:2002we,Balantekin:2004ug}.
In fact, collective flavor transformation in the supernova
environment previously was thought to be of the
synchronized, collective form \cite{Pastor:2002we}.
However, \cite{Duan:2007fw} shows that the collective flavor
transformation observed in this simulation is not
synchronization, but rather a neutrino-background-enhanced MSW-like
flavor transformation.
This phenomenon occurs where a neutrino with an energy representative
of the neutrino-antineutrino gas would encounter a 
conventional MSW resonance. Because the representative energy
of a dense neutrino-antineutrino gas is usually smaller
than the average energy of each neutrino species, this type of collective
flavor oscillation can occur deeper 
in the supernova envelope than
one would expect for pure matter-driven MSW.

In our simulations we find that neutrinos and
antineutrinos emitted in different directions behave differently.
Of course, this is in direct contradiction to the single-angle
approximation.
In our normal neutrino mass hierarchy simulations,
large-scale flavor transformation
 is initiated by neutrinos and antineutrinos in the most
tangentially-propagating
angle bins (i.e., those with low values of $\cos\vartheta_0$). As
the radius increases, we can see from the 
\href{http://simulations4snu.googlepages.com/late_time_2flavor_nh.html}{%
full simulation}  that flavor conversion
spreads out to other angle bins, eventually causing 
significant flavor transformation of both neutrinos and 
antineutrinos for nearly all energies and almost 
all trajectory directions. 

The spreading of significant flavor transformation from tangential to
radial directions
has its origin in the structure of the weak current.
More tangentially-directed neutrinos or antineutrinos which
forward-scatter on the 
more radially-directed background neutrinos and antineutrinos bring
larger flavor diagonal and off-diagonal potentials 
to the full self-coupling Hamiltonian $\hat{H}_{\nu\nu}$. This is a
consequence of the
intersection angle dependence of the forward neutrino-neutrino
scattering amplitude \cite{Fuller:1987aa,Fuller:2005ae,Duan:2006an,Duan:2006jv}.

As we move 
out to even larger radius in these simulations, 
neutrinos and antineutrinos experience collective bipolar oscillations. 
This behavior is evident in the
\href{http://simulations4snu.googlepages.com/late_time_2flavor_nh.html}{%
full simulation}. At large enough radius the neutrino background becomes
ineffective
in influencing flavor transformation. The subsequent neutrino and
antineutrino
flavor evolution is akin to MSW or vacuum neutrino oscillations.
However, as will be discussed below,
the breakdown of collective flavor transformation can leave an important
``fossil'' imprint of the nonlinear neutrino background.

\begin{figure}[t]
\href{http://simulations4snu.googlepages.com/late_time_2flavor_ih.html}{%
\includegraphics*[origin=bl,width=\textwidth,keepaspectratio]%
{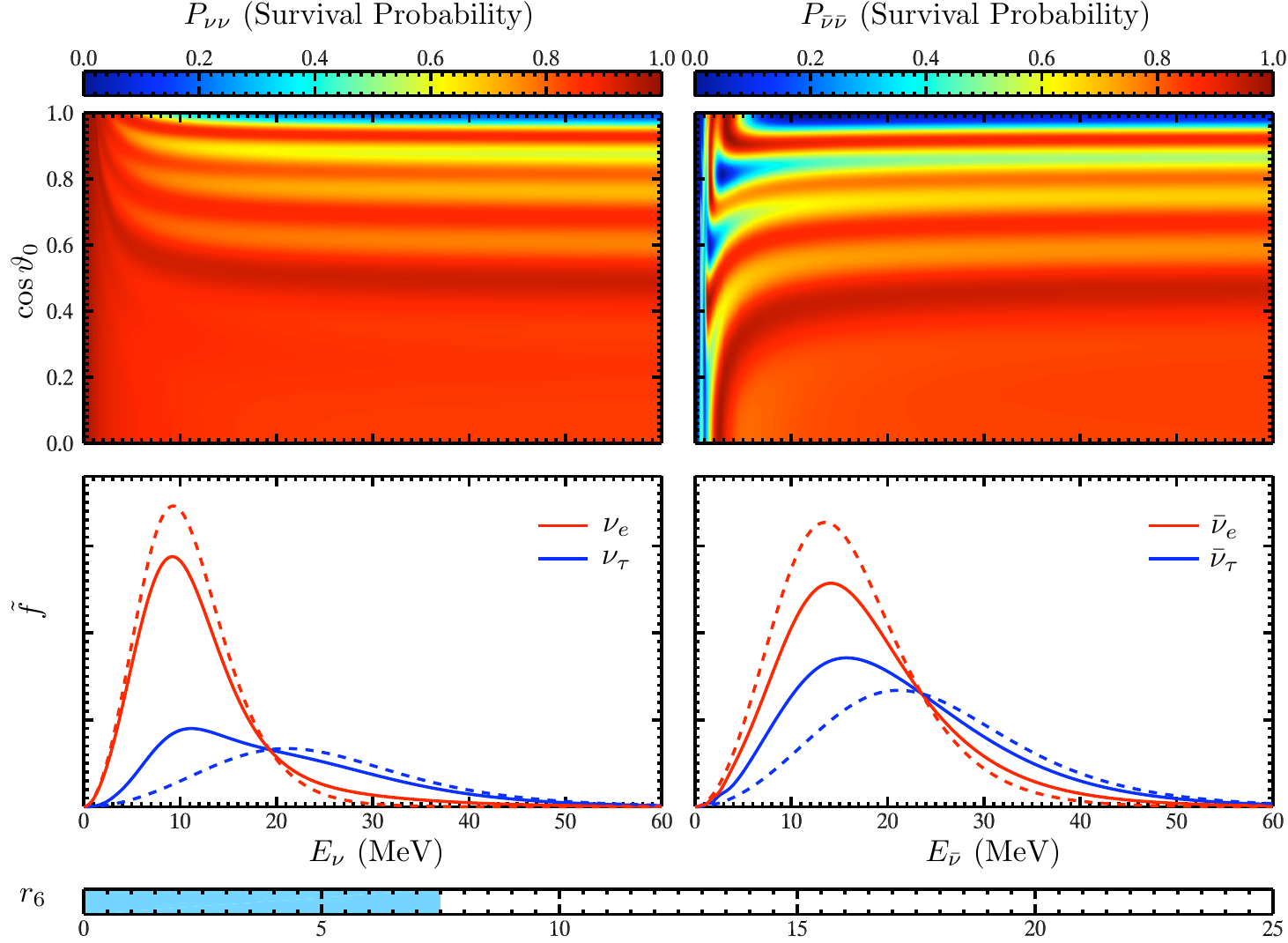}}
\caption{\label{fig:inverted}%
Neutrino and antineutrino flavor transformation results 
for the inverted neutrino mass hierarchy case
but now with full neutrino self-coupling 
and trajectory coupling.
Setups and definitions are the same as in Figure \ref{fig:normalMSW}.
Here we take take the luminosities of all neutrino species to be
$L_\nu=10^{51}$ erg/s.
This plot corresponds to the snapshot 
at $r_6\simeq7.5$ in the 
\href{http://simulations4snu.googlepages.com/late_time_2flavor_ih.html}%
{full simulation}.}
\end{figure}

In the inverted mass hierarchy, neutrino and antineutrino flavor
evolution with self-coupling
is different from the behavior in both the matter-driven MSW
case and the normal neutrino mass hierarchy self-coupled case.
Figure \ref{fig:inverted} shows a snapshot ($r\simeq75$ km) of
one of our simulations for the
inverted neutrino mass hierarchy case with the full
neutrino self-coupling Hamiltonian
($L_{\nu_e}=L_{\bar\nu_e}=L_{\nu_\tau}=L_{\bar\nu_\tau}=10^{51}$ erg/s).
A \href{http://simulations4snu.googlepages.com/late_time_2flavor_ih.html}%
{movie} presenting the full simulation is available for downloading.
Like the normal mass hierarchy case, both neutrinos and antineutrinos
can experience large-scale, and simultaneous flavor conversion
over broad ranges of energies and for nearly all emission angles.
However, in contrast to
the normal mass hierarchy behavior, 
neutrinos and antineutrinos in the inverted mass hierarchy case do not
experience MSW-like flavor transformation, but rather enter the
bipolar oscillation mode directly
at low radius. Subsequently, as the radius increases, more and more
horizontal fringes
appear in the survival probability $P_{\nu\nu}(\cos\vartheta_0,E_\nu)$
and $P_{\bar\nu\bar\nu}(\cos\vartheta_0,E_\nu)$.
These horizontal fringes also appear, although to a less ``violent'' extent, 
in the normal mass hierarchy case. This phenomenon is associated
with multi-angle effects. 
The one-by-one excitation of multipoles in $\cos\vartheta_0$ 
so evident in our 
\href{http://simulations4snu.googlepages.com/late_time_2flavor_ih.html}%
{full simulation} is explained by the
``kinematic decoherence'' of bipolar oscillations \cite{Raffelt:2007yz}.

Neutrino and antineutrino flavor transformation in the inverted neutrino
mass hierarchy case sets in at a radius $r_\mathrm{X}$ 
where the total neutrino flux drops below a
critical value. This can occur closer 
to the neutron star than  MSW-like transformation
in the normal neutrino mass hierarchy case.
Because $r_\mathrm{X}$ is solely  determined by neutrino
fluxes and energy spectra in this case
\cite{Duan:2005cp},
neutrinos and antineutrinos can experience
large-scale flavor transformation even in earlier epochs,
where the matter density is much higher.
Indeed, a simple classical pendulum analogy shows that
the inverted mass hierarchy gives rise
to an inherently unstable neutrino flavor field, like a pencil standing
on its tip \cite{Hannestad:2006nj,Duan:2007mv}.  

Although the single-angle approximation may not be accurate in general, the
calculations
based on this assumption do capture some if not most of the {\it
qualitative} features
present in multi-angle calculations. This has been shown in our
simulations
\cite{Duan:2006an,Duan:2006jv} as well as in the computations carried
out
by other groups \cite{EstebanPretel:2007ec,Fogli:2007bk}.

One of the most interesting behaviors seen in both the
single-angle
and multi-angle calculations is
stepwise spectral swapping (also known as the spectral split)
\cite{Duan:2006an,Duan:2006jv}. 
This is where, e.g., $\nu_e$ with energies larger or smaller
than some transition
energy $E^\mathrm{s}$ swap energy spectra and fluxes with neutrinos of
another flavor, say $\nu_\tau$. The swapping feature is strikingly
obvious in, e.g., the later part (larger radius part) of 
\href{http://simulations4snu.googlepages.com/late_time_2flavor_nh.html}{%
the simulation} that
corresponds to Figures \ref{fig:normal}.
Note the vertical (i.e., for all
$\vartheta_0$) line of demarcation between near-zero survival
probability for neutrinos with $E_\nu < E^\mathrm{s} \simeq 10\,{\rm
MeV}$ and significant survival probability for more energetic neutrinos.
``Swapping''  is an apt description of this phenomenon
in the sense that there is nearly complete conversion 
of neutrino flavors across a broad range of neutrino energy $E_\nu$.
Whether the swapping of spectra occurs for values of $E_\nu$ above
or below $E^\mathrm{s}$ depends on the neutrino mass hierarchy. 
In the normal (inverted)
neutrino mass hierarchy the swap occurs 
for neutrinos with energies satisfying
$E_\nu < E^\mathrm{s}$ ($E_\nu >
E^\mathrm{s}$). An explanation of this phenomenon 
in the single-angle approximation context has been given by
\cite{Duan:2006an,Duan:2007mv,Raffelt:2007cb,Duan:2007fw,Raffelt:2007xt}. 

In our simulations with the normal neutrino mass hierarchy, the swap
energy $E^\mathrm{s}$ decreases as the effective mixing angle
$\theta\,(\simeq\theta_{13})$ is decreased. 
The situation in the inverted neutrino mass hierarchy
is different. As alluded to above, there is ``instability''
in flavor transformation in the inverted mass scheme. Indeed, our
simulations suggest that a swap can be obtained in this scheme 
even for extremely small values of $\theta$ \cite{Duan:2007bt}.

The direct
detection of stepwise swapping in the energy spectra of
supernova neutrinos could provide a unique way to determine
the neutrino mass hierarchy, even if $\theta_{13}$ is too small
to be measured in conventional 
terrestrial neutrino oscillation experiments 
\cite{Duan:2007bt}. This is an important result that could change future
supernova neutrino detection strategies. 
The swap energy is relatively small, $E^\mathrm{s}\sim 10\,{\rm MeV}$,
on the order of solar neutrino energies.
These facts may suggest that supernova
neutrino detection schemes should
focus on low energy, low flux $\nu_e$'s at a few seconds
post-core-bounce.

We have also run 3-neutrino (i.e., $3\times 3$) flavor 
mixing simulations, where all three neutrino flavors 
are followed simultaneously and self-consistently.
Given the wealth of unexpected
phenomena revealed by the nonlinear $2\times 2$ mixing
calculations,
it would be prudent to investigate whether going to a full $3\times 3$
mixing scheme might change things yet again. Sometimes
the full 3-neutrino mixing problem should be be reducible to two
separate 2-flavor problems 
\cite{Kuo:1986sk,Dighe:1999bi,Balantekin:1999dx,Caldwell:1999zk}. 
However, this is not always possible.

We have run both single-angle
and multi-angle tests which employ the full 3-flavor mixing scheme
for one particular case: the neutronization burst from
an O-Ne-Mg core-collapse supernova \cite{Duan:2007sh,Duan:2008zb}.
The neutronization neutrino burst
occurs when the supernova bounce shock comes through the neutrino sphere
at $\sim 10\,{\rm ms}$ post-core-bounce. O-Ne-Mg core collapse
supernovae have as progenitors
stars with masses in the range $8$ to $12\,{\rm M}_\odot$. These
are very interesting from a neutrino physics standpoint because
there is little matter overburden for the shock to transit and, as a
result,
the neutrino background can become important very early on
even in the normal neutrino mass hierarchy case.
For this case, in the normal neutrino mass hierarchy, our $3\times 3$
numerical calculation
shows two stepwise spectral swaps for neutrinos in
the vacuum mass basis, one on the top of the other. 
This result seems to suggest that, at least for this particular case,
the full 3-flavor mixing problem is indeed reducible to two
separate 2-flavor mixing schemes. The result for the inverted
mass hierarchy case is similar, except that one of the
spectral swaps is forbidden because of the conservation of
a ``lepton number'' \cite{Hannestad:2006nj,Raffelt:2007cb}.

\section{Frontiers%
\label{sec:frontiers}}

Though there has been dramatic recent progress in understanding 
the evolution of the neutrino flavor field in the supernova environment, 
there are many outstanding and unresolved issues.
As recent multi-dimensional hydrodynamics simulations have shown 
(e.g., \cite{Kifonidis:2005yj,Blondin:2006yw}),
the actual supernova matter density could be quite inhomogeneous. 
In addition, it is possible that the neutrino flux and angular 
distributions could deviate significantly from
the idealized spherically symmetric models employed in our numerical 
simulations. Nevertheless
we believe that our spherically symmetric coherent regime calculations 
at the very least capture the qualitative features of the collective 
flavor oscillation modes discussed above. But a road map for future
 work should include a plan for checking this assertion.


One of the most significant limitations in current multi-angle simulations 
is the neutrino bulb model employed in these
calculations. In the neutrino bulb model, neutrinos are assumed
to be emitted isotropically from an infinitely thin layer (i.e., the
neutrino sphere) on the surface of a spherically symmetric black-body
(the neutron star). In addition, in current neutrino flavor 
transformation simulations the matter
profile outside the neutron star is taken to depend only on the radius. 

This model roughly corresponds to the 1D
supernova simulations, but with nowhere
near the sophistication  of those calculations. 
For example, it is well known that
neutrinos with different flavors and with different energies
experience different opacities
and, therefore, decouple from the matter fields at different radii. 
Even in the context of 1D supernova models,
the initial neutrino energy spectra will show deviations from black-body form.
Moreover, the neutrino sphere has a finite thickness. 
This will alter the relationship between
length parameters and emission angles on neutrino trajectories.

Recent sophisticated supernova simulations suggest that
3D features might play significant
roles in how massive stars eventually collapse. Needless to say,
neutrino emission would not be isotropic in these simulations, and
the matter profile obviously would not be spherically symmetric either. 
In order for  \textit{FLAT} 
to treat the 3D supernova features 
in neutrino flavor evolution, a new \texttt{NBGroup\_EM} module would be
necessary. This module would have to model correctly the 
physics and geometry of the problem.
\textit{BULB} would have to be modified in corresponding fashion.
Of course, actually implementing such modifications in either 
code would significantly enlarge the dimensionality
of the problem space and, once achieved, could increase
the computation time by several orders of magnitude.  

Flavor evolution of neutrinos emitted from the accretion disk
around a massive black hole is an interesting example of non-sphericity.
Even the simplest accretion disk model has a more complicated 
geometry than the neutrino bulb. In principle, this problem
could be approached in a manner similar to the 1D to 3D upgrade, 
i.e., by building
a new \texttt{NBGroup\_EM} module to treat the special geometric structure
of the accretion disk.

Yet another interesting problem is the flavor evolution of
the neutrinos and antineutrinos in the early universe. This was the
environment where the first full nonlinear numerical simulations of
neutrino flavor evolution were performed
\cite{Samuel:1993uw,Kostelecky:1993ys,Kostelecky:1993yt,Kostelecky:1993dm,Kostelecky:1994dt,Samuel:1995ri,Kostelecky:1996bs}.
These early explorations revealed several interesting features
of flavor evolution in dense neutrino gases. These features included, 
for example, synchronized
and bipolar collective neutrino oscillations. This work has had 
important implications 
for models of lepton-degenerate cosmologies (see, e.g., 
\cite{Dolgov:2002ab,Wong:2002fa,Abazajian:2002qx}). Because
the early universe and supernovae have, in some sense, 
similar physical environments, the interesting phenomena recently
discovered in the supernova case might also occur in the early universe.
It could be worthwhile to re-investigate neutrino flavor
evolution in the early universe too, for example, to see if bipolar 
neutrino oscillations
might have consequences for models for lepton number
affected nucleosynthesis. 

Following neutrino and antineutrino flavor evolution 
in astrophysical environments
is a difficult problem. However, astrophysicists are 
compelled to solve this problem because
of the potential leverage neutrino flavor transformation
 has in supernova and early universe physics, and because 
neutrino flavor mixing is an experimental reality. 
We believe that the solution of the astrophysical neutrino
 flavor transformation problem could usher in a new era where 
there is a synergistic coupling of terrestrial laboratory-based 
neutrino physics and astrophysical observations.

\ack
This work was supported in part by
NSF grant PHY-04-00359, a LANL/CARE grant
and TSI collaboration's
DOE SciDAC grant at UCSD,
DOE grant DE-FG02-00ER41132 at INT,
and by the DOE Office of Nuclear Physics, the LDRD Program
and Open Supercomputing at LANL, and an IGPP/LANL mini-grant at UCSD.
This work was also supported in part by NERSC
 through TSI collaboration using Bassi, SDSC
through AAP using DataStar,
and LCF at NCCS using Jaguar.

\section*{References}
\bibliographystyle{iopart-num}
\bibliography{nubulbmovie}

\end{document}